\documentclass[]{interact}

\usepackage{color}
\usepackage{epstopdf}
\usepackage[caption=false]{subfig}

\usepackage[numbers,sort&compress]{natbib}
\bibpunct[, ]{[}{]}{,}{n}{,}{,}
\makeatletter
\def\NAT@def@citea{\def\@citea{\NAT@separator}}
\makeatother

\theoremstyle{plain}
\newtheorem{theorem}{Theorem}[section]

\theoremstyle{definition}

\theoremstyle{remark}
\newtheorem{remark}{Remark}

\begin{document}


\title{Testing semiparametric model--equivalence hypotheses based on the characteristic function}

\author{
\name{Feifei Chen\textsuperscript{a}, Simos G. Meintanis\textsuperscript{b,c,$\ast$}\thanks{\textsuperscript{$\ast$}On sabbatical leave from the University of Athens.}, and Lixing Zhu\textsuperscript{a,$\dagger$}\thanks{\textsuperscript{$\dagger$}Lixing Zhu is the corresponding author. E-mail: lzhu@hkbu.edu.hk}}
\affil{\textsuperscript{a}Center for Statistics and Data Science, Beijing Normal University, Zhuhai, China; \\
\textsuperscript{b}Department of Economics, National and Kapodistrian University of Athens, Athens, Greece; \\
\textsuperscript{c}Pure and Applied Analytics, North--West University, Potchefstroom, South Africa}
}

\maketitle

\begin{abstract}
We propose three test criteria each of which is appropriate for testing, respectively, the equivalence hypotheses of symmetry, of homogeneity, and of independence, with multivariate data.
All quantities have the common feature of involving weighted--type distances between characteristic functions and are convenient from the computational point of view if the weight function is properly chosen.
The asymptotic behavior of the tests under the null and alternative hypotheses is investigated.
Numerical studies and a real--data application are conducted in order to examine the performance of the criteria in finite samples.
\end{abstract}


\section{Introduction}

In a parametric framework the term ``precise hypothesis'' is mostly used in bioequivalence when a {\it{point}} null hypothesis $\vartheta=\vartheta_0$ is being tested about a given parameter $\vartheta$.
The issue of whether a point null hypothesis such as this is reasonable or if a less precise hypothesis would be more appropriate has a long history and is being strongly debated often in connection to the so--called ``Lindley's Paradox''.
This paradox stated in rough terms says that as the sample size increases any frequentist approach tends to lean towards the alternative while from the Bayesian perspective the data will become even more inclined towards the null hypothesis; see for instance \cite{Berger1987, Sprenger2013}.
In view of this debate some researchers have opted towards reformulation of the problem as one of testing,  for some fixed $\Delta>0$, the 
neighborhood--type null hypothesis  $|\vartheta-\vartheta_0|\leq \Delta$, whereby as opposed to a point null, the value of the parameter is only approximately specified under the null.  Rejection then of the null implies that
$|\vartheta-\vartheta_0|>\Delta$, i.e. that $\vartheta$ is at least at a distance $\Delta$ from $\vartheta_0$.
(Note that $\Delta$  can be chosen as a shrinking neighborhood of the sample size, analogously to the common practice of power analysis by means of contiguous alternatives; \cite[see][\S49 and \S55]{Borovkov1998}).

Recently there is even a tendency to interchange the null and alternative hypotheses,
with the resulting, so--called \textit{equivalence} hypothesis formulated as
$$
  {\cal {H}}_0: |\vartheta-\vartheta_0| \geq \Delta \ \ \ \mbox{versus} \ \ \ {\cal {H}}_1: |\vartheta-\vartheta_0|< \Delta.
$$
Thus rejection of the null indicates that the parameter $\vartheta$ is ``$\Delta$--close'' to $\vartheta_0$.
This formulation, which has become standard by now, also has its roots in the area of bioequivalence problems which are distinguished from conventional testing problems. 
 For example, in the specific case of testing for bioequivalence of a new treatment against an established treatment, we are faced with a parameter $\vartheta$ measuring (potential) improvement from $\vartheta_0$ which corresponds to an established treatment. Then a threshold, say $\Delta$,  is set on this improvement so that rejection of the null hypothesis $|\vartheta-\vartheta_0| \geq \Delta$ in favour of the alternative $|\vartheta-\vartheta_0|< \Delta$ implies bioequivalence of the two treatments, meaning that the effect of the new treatment, even if positive, is rather minuscule.
In this restricted context of inference about a parameter there is considerable literature for which the monograph of \cite{Wellek2010} is a good starting point.

On the other hand the problem of testing ``model--equivalence'' (or ``neighborhood--of--model validation") has only recently been considered in a more general context, beyond that of testing about a single parameter.

Generally speaking, the context of testing model--equivalence is that of test a hypothesis of the form
${\cal {H}}_0: d \geq \Delta \ \mbox{against} \ {\cal {H}}_1: d < \Delta, $
where $d$ is a real-valued parameter that measures the distance between two models.
For example, \cite{Dette2018} constructed tests for the equivalence of two regression curves based on the $L_2$ distance and maximal deviation distance between two parametric regression models.
We also refer to \cite{Dette1998, Baringhaus2017, Dette2018, Henze2020, Dette2021, Mollenhoff2022}, for testing model--equivalence in various settings.

In this paper we will restate the three major semiparametric problems of distributional homogeneity (or the two--sample problem), symmetry and independence in the form of model--equivalence hypotheses testing. 
That is, we will develop statistical methodology for testing homogeneity--equivalence, symmetry--equivalence, and independence--equivalence, respectively.
We now shortly discuss the importance and applicability of the independence--equivalence test.
(The discussion for the  other two equivalence tests is analogous).
In this connection, it is well known that measuring and testing dependence is  a fundamental task in statistical inference and data analysis, for instance in gene selection and causal inference.
Classical independence tests however typically only consider the exact (or precise) independence corresponding to the point null  hypothesis ${\cal {H}}_0: d=0$, which as already discussed in the first paragraph is prone to Lindley's Paradox, at least in the current era of massive data.
In addition, with many applications such as detecting the dependence between treatment effect and covariates in causal inference, one may be interested in detecting dependence only beyond a certain threshold of given size, with this size of dependence of interest carefully chosen on the basis of the underlying context.
Thus practitioners may wish to detect dependence beyond a certain fixed size of interest by specifying the threshold $\Delta$ in the independence--equivalence test, the specifics of which will be studied in detail in this paper and will be illustrated by a real--data application in Section 5.

In the following, we will use the word ``model'' as synonymous with the symmetry of a given random variable, while for two random variables or pairs of variables, to denote distributional homogeneity or independence of these variables, respectively.
Specifically for a given model, say $\mathcal M$, and an arbitrary random vector $Z$ we propose  the (model--specific) population measure
\begin{eqnarray} \label{model}
  \Delta^{(\mathcal M)}_{w,Z}=\int \left|D^{(\mathcal M)}_Z(t)\right|^2 w(t){\rm{d}}t,
\end{eqnarray}
which satisfies $\Delta^{(\mathcal M)}_{w,Z}=0$, if and only if the law of random vector $Z$ lies within model $\mathcal M$, where $D^{(\mathcal M)}_Z(\cdot)$ is a pointwise distance from the null hypothesis, and $w>0$ denotes a weight function to be further discussed along the paper. The composition of the vector $Z$ will vary depending on the context, consisting of a single vector for the case of testing for symmetry, or of two independent random vectors of equal dimensions in the case of  homogeneity, or consisting of a pair of random vectors of potentially different dimensions in the case of independence testing.
In this framework we consider the model--equivalence problem of testing the null hypothesis
\begin{eqnarray} \label{null}
  {\cal {H}}^{(\mathcal M)}_0: \Delta^{(\mathcal M)}_{w,Z}\geq \Delta,
\end{eqnarray}
against the alternative
\begin{eqnarray} \label{alt}
  {\cal {H}}^{(\mathcal M)}_1: \Delta^{(\mathcal M)}_{w,Z}< \Delta,
\end{eqnarray}
so that if ${\cal {H}}^{(\mathcal M)}_0$ is false this will imply that the law of $Z$ lies within a neighborhood of model $\cal M$ of fixed length $\Delta > 0$. (Formally, the null hypothesis ${\cal {H}}^{(\mathcal M)}_0$ as well as the alternative should be indexed by $\Delta$, but we will suppress this dependence for simplicity).

The rest of the paper unfolds as follows.
In Section \ref{sec2} we introduce the null and alternative hypotheses, propose an appropriate population measure $\Delta^{(\mathcal M)}_{w,Z}$ in each case, and formulate our testing approach in general terms.
In Section \ref{sec3}, the three test criteria are specified along with the weight function $w(\cdot)$ and asymptotic results are presented.
The finite--sample properties of the test criteria are studied in Section \ref{sec4} by Monte Carlo experiments, while in Section \ref{sec5} we illustrate the applicability of our methods by means of a real--data application.
The paper concludes in Section \ref{sec6} with discussion and outlook. Technical proofs are postponed to the Appendix.

\section{Population measures, computations and test criteria} \label{sec2}
\subsection{Population measures} \label{sec21}
We will consider population measures that are formulated as $L_2$--type distances, and measure discrepancy between symmetry and asymmetry of a given law, while for pairs of distributions they express distance between these distributions, or distance between independence and arbitrary modes of dependence of corresponding random variables. These population measures are formulated in terms of  characteristic functions (CFs).
Specifically suppose that $Z:=\{X, Y\}$ where $X\in\mathbb R^p$ and $Y\in\mathbb R^p$ are two independent random vectors, and write $\varphi_{X}(\cdot)$ and $\varphi_{Y}(\cdot)$ for the corresponding CFs.
Then we define the CF--distance between $X$ and $Y$ as 
\begin{eqnarray} \label{CFH}
  \Delta^{(\mathcal H)}_{w,Z} &=& \int_{\mathbb R^p} \left|\varphi_X(t)-\varphi_Y(t)\right|^2 w(t){\rm{d}}t.
\end{eqnarray}
We wish to test the null hypothesis figuring in \eqref{null} against the alternative hypothesis \eqref{alt} by means of an empirical counterpart of $\Delta^{(\mathcal H)}_{w,Z}$.
In this setting rejection of ${\cal {H}}^{(\mathcal M)}_0$ would imply that the laws of the two random variables are equivalent, or $\Delta$--close, in terms of CFs.

Note that the law of a random vector $X\in\mathbb R^p, \ p\geq 1$, is symmetric around the origin if and only if ${\tt {Im}}(\varphi_X(t))=0, \forall t\in\mathbb R^p$, where ${\tt Im}(z)$ denotes the imaginary part of a complex number $z$.
Then the next population measure corresponds to testing symmetry is defined by setting $Z:= X$ in \eqref{model}, as
\begin{eqnarray} \label{CFS}
  \Delta^{(\mathcal S)}_{w,Z} &=& \int_{\mathbb R^p} \left\{ {\tt {Im}}(\varphi_X(t)) \right\}^2 w(t){\rm{d}}t.
\end{eqnarray}
Again rejection of the null hypothesis would imply that the law of $X$ belongs to the equivalence class of all centrally symmetric distributions, in the sense of being within  a (CF--based) neighborhood of symmetry of size $\Delta$.

Our last population measure  will be used to measure the significance of dependence between a pair of random vectors $X\in \mathbb R^p$ and $Y \in \mathbb R^q$.
To this end set $Z:=(X^\top,Y^\top)^\top$, write $\varphi_{X,Y}(\cdot,\cdot)$ for the CF of $Z$ and define
\begin{eqnarray} \label{CFI}
  \Delta^{(\mathcal I)}_{w,Z} &=& \int_{\mathbb R^{p+q}} \left|\varphi_{X,Y}(t_1,t_2)-\varphi_X(t_1) \varphi_Y(t_2)
  \right|^2 w(t_1,t_2){\rm{d}}t_1 {\rm{d}}t_2, \end{eqnarray} where $w(t_1,t_2)=w_p(t_1) w_q(t_2)$, with each $w_p(\cdot)$ and $w_q(\cdot)$ being a weight function in the indicated dimension.


\subsection{Specific computations} \label{sec22}
We note that CF--based population measures such as these figuring in 
\eqref{CFH}--\eqref{CFI} may be further manipulated to express distance in terms of more conventional quantities. The readers are referred to
\cite[see][\S 3]{Chen2019} for such interpretations. For analogous interpretations based on moments see \cite{Szekely2007, Szekely2013}. By way of example we consider here the homogeneity distance. In this connection further below we will assume that the weight function is a density of a spherical distribution but for our present purposes also assume that this density  may be written as $w(t)=c \: \varphi^2_V(t)$, where $c>0$ is a global constant and $\varphi_V(t)$ denotes the CF of a random vector $V$ having a symmetric around zero distribution. A representation like this is possible for a few distributions such as the $p$--dimensional Kotz-type distribution with density $f(x)=c \exp(-2\|x\|^{\gamma}), 0<\gamma\leq 2$, and the generalized spherical Laplace distribution 
with density $f(x)=c(1+\|x\|^2)^{-(p+1)/2}$.
Here $\|\cdot\|$ denotes the Euclidean norm.
In the Kotz-case, $V$ corresponds to a spherical stable distribution with CF $\exp(-\|t\|^{\gamma})$, while in the latter case to a generalized spherical Laplace distribution with CF $(1+\|t\|^2)^{-(p+1)/4}$.
We refer to \cite{Nadarajah2003, Kozubowski2013} for the Kotz-type distribution and the Laplace distribution, respectively.

With such a weight function and by using Parseval's identity we have from \eqref{CFH}, 
\begin{eqnarray} \label{parseval} \nonumber
  \Delta^{(\mathcal H)}_{w,Z} 
&=& \int_{\mathbb R^p}|\varphi_{X}(t)-\varphi_{Y}(t)|^2 c \: \varphi^2_V(t){\rm{d}}t \\ \nonumber
&=& c\int_{\mathbb R^p}|\varphi_{X\ast V}(t)-\varphi_{Y\ast V}(t)|^2{\rm{d}}t \\ \nonumber
&=& c(2\pi)^p \int_{\mathbb R^p}(f_{X\ast V}(x)-f_{Y\ast V}(x))^2{\rm{d}}x,
\end{eqnarray}
where $\ast$ denotes convolution. Thus apart from a constant, the CF-based homogeneity distance between a pair of distributions is equal to a distance associated with the densities of these distributions, whereby the weight function $w(t)$ acts on these densities as convolution with the density corresponding to the CF $\varphi_V(t)$ involved in $w(t)$.

Another interpretation in terms of moments is provided by the energy distance measure of \cite{Szekely2013},
\begin{eqnarray} \label{szekely}
{\mathcal {E}}_\gamma(X,Y) &=& \mathbb E\left[2 \|X-Y\|^\gamma - \|X-X_1\|^\gamma - \|Y-Y_1\|^\gamma \right],
\end{eqnarray}
which may be obtained from the homogeneity distance in \eqref{CFH} if $w(t)$ is proportional to $\|t\|^{-(p+\gamma)}$, with $\gamma\in(0,2)$, under the assumption  $\mathbb E\left[\|X\|^\gamma\right] < \infty$, and $\mathbb E\left[\|Y\|^\gamma\right] < \infty$.

Passing to the problem of independence, we note that correlation, which is arguably the most straightforward measure of independence, has been shown by \cite{Szekely2007} to be directly related to distance covariance, which again may be obtained from the independence distance in \eqref{CFI} by setting the weight function proportional to $\|t_1\|^{-(p+\gamma)}\|t_2\|^{-(q+\gamma)}$, under certain moment assumptions. We will not repeat the arguments here but simply compute the independence measure for a bivariate zero-mean Gaussian distribution with unit component variances and correlation equal to $\varrho$, say $\Delta_{\varrho,G}^{(\mathcal I)}$. Specifically with weight function $w(t)=\exp(-t_1^2-t_2^2)$ we obtain from \eqref{CFI},
\begin{eqnarray*}
\Delta_{\varrho,G}^{(\mathcal I)}
&=& \int_{\mathbb R^2} \left|\exp\left(-\frac{t_1^2+t_2^2+2 \varrho t_1 t_2}{2}\right) - \exp\left(-\frac{t^2_1+t^2_2}{2}\right)\right|^2 \exp\left(-t^2_1-t^2_2\right){\rm{d}}t_1{\rm{d}}t_2 \\
&=& \pi\left(\frac{1}{2}+\frac{1}{\sqrt{4-\varrho^2}}-\frac{4}{\sqrt{16-\varrho^2}}\right),
\end{eqnarray*}
which is increasing with $|\varrho|$ and vanishes at $\varrho=0$. It is also immediate from \eqref{CFH} that  $\Delta_{\varrho,G}^{(\mathcal I)}$ coincides with the population homogeneity distance $\Delta_{\varrho,G}^{(\mathcal H)}$ between the aforementioned bivariate Gaussian distribution and a corresponding bivariate zero-mean Gaussian distribution with independent components, i.e. with $\varrho=0$. In both cases the parameter of interest is equal to $\varrho$, with smaller (larger) $|\varrho|$ implying smaller (larger) deviation from the corresponding exact property of independence or homogeneity, and thus values of $\Delta_{\varrho,G}^{(\mathcal I)}$ and  $\Delta_{\varrho,G}^{(\mathcal H)}$ admit a simple interpretation in terms of the corresponding parameter of interest.

As a further example in arbitrary dimension $p\geq 1$, we consider a two-component equal mixture of a standard Gaussian distribution with the same Gaussian distribution shifted by $\delta \in \mathbb R^p$. For weight function $\exp(-\|t\|^2)$ we compute from \eqref{CFS} the resulting symmetry distance for this mixture distribution as
\[
\Delta_{\delta,G}^{(\mathcal S)} 
= \int_{\mathbb R^p} \frac{1-\cos(2 \delta^\top t)}{8} \exp\left(-2 \|t\|^2\right) {\rm{d}}t
= \frac{\pi^{\frac{p}{2}}}{2^{\frac{p}{2}+3}}\left(1-\exp\left(-\frac{\|\delta\|^2}{2}\right)\right),
\]
which is again an easily interpretable function of the parameter of interest $\delta$. The corresponding homogeneity distance between this mixture  and a standard Gaussian distribution may also be computed as
\[
\Delta_{\delta,G}^{(\mathcal H)}= \frac{\pi^{\frac{p}{2}}}{2^{\frac{p}{2}+1}}\left(1-\exp\left(-\frac{\|\delta\|^2}{8}\right)\right), \]
and the same observations apply. Thus, and although these examples are clearly only indicative, it appears that when a certain property (symmetry, homogeneity or independence) is expressed via a given parameter, the corresponding distance measure is a function (perhaps complicated) having properties that are in line with intuition regarding the role of this parameter. We close this paragraph by stressing that the range of possible values of $\Delta_{w,Z}^{(\mathcal M)}$, when placed in the appropriate context,  is important in order  to acquire  a feeling about meaningful values of the threshold parameter $\Delta$, which is again heavily context-dependent; see for instance \cite[\S 1.7]{Wellek2010}. Yet this range is not always as straightforward to compute as in the previous examples, and in fact may require high-dimensional numerical integration.

\subsection{Test criteria} \label{sec23}
Now assume that $\Delta^{(\mathcal M)}_{n,w}$ is an empirical counterpart of $\Delta^{(\mathcal M)}_{w,Z}$ resulting from a sample of size $n$ on $Z$,  such that
\begin{eqnarray} \label{CLT}
  \sqrt{n}\left(\Delta^{(\mathcal M)}_{n,w}-\Delta^{(\mathcal M)}_{w,Z}\right) \stackrel{{\cal D}}{\longrightarrow} \mathcal N\left(0,\sigma_w^{2(\mathcal M)}\right)
\end{eqnarray}
when $\Delta^{(\mathcal M)}_{w,Z}>0$, and
\begin{eqnarray} \label{asy}
   \sqrt{n} \Delta^{(\mathcal M)}_{n,w} \stackrel{\mathbb  P}{\longrightarrow} 0
\end{eqnarray}
when $\Delta^{(\cal{M})}_{w,Z}= 0$.
Here $\stackrel{{\cal D}}{\longrightarrow}$ and $\stackrel{\mathbb  P}{\longrightarrow}$ signify convergence in distribution and in probability, respectively.
Suppose further that the limit variance $\sigma_w^{2(\mathcal M)}$ (again we economize on notation by suppressing dependence of $\sigma_w^{2(\mathcal M)}$ on $Z$) can  be estimated by, say, $\sigma_{n,w}^{2(\mathcal M)}$, such that
\begin{eqnarray} \label{varn}
   \sigma_{n,w}^{2(\mathcal M)} \stackrel{\mathbb  P}{\longrightarrow} \sigma_w^{2(\mathcal M)}.
\end{eqnarray}
This limit variance $\sigma_w^{2(\mathcal M)}$  as well as its consistent estimator $\sigma_{n,w}^{2(\mathcal M)}$ will be discussed in detail in Section \ref{sec3}.

On the basis of the preceding discussion and for some fixed threshold $\Delta>0$, we suggest the test defined by the critical region
\begin{eqnarray} \label{CR}
  \Delta^{(\mathcal M)}_{n,w} \leq \Delta + \frac{\sigma_{n,w}^{(\mathcal M)}}{\sqrt{n}} z_\alpha,
\end{eqnarray}
where $\sigma_{n,w}^{(\mathcal M)}$ is the positive square root of $\sigma_{n,w}^{2(\mathcal M)}$, and $z_\alpha$ is the $\alpha$ quantile of the standard normal distribution.
Clearly rejection of ${\cal {H}}^{(\mathcal M)}_0$ implies model--equivalence, or more precisely that the law of our data is $\Delta$--close to model $\mathcal M$ under the corresponding distance measure $\Delta^{(\mathcal M)}_{w,Z}$.
In fact it readily follows that the criterion with critical region defined by \eqref{CR} leads to  an asymptotic size--$\alpha$ test, which is consistent against fixed alternatives.
Specifically for each law of $Z$ with corresponding population measure $\Delta^{(\mathcal M)}_{w,Z}>0$,
if \eqref{CLT} and \eqref{varn} hold true, we have for the probability of rejection of the null hypothesis in \eqref{null} that
\begin{eqnarray} \label{reject}
 & & \lim_{n\to\infty} \mathbb P \left( {\rm{Reject}} \ {\cal {H}}_0^{(\mathcal M)} ~\big|~ \Delta^{(\mathcal M)}_{w,Z} > 0 \right) \nonumber \\
 &=& \lim_{n\to\infty} \mathbb P\left(\frac{\sqrt{n}\left( \Delta^{(\mathcal M)}_{n,w} - \Delta \right)}{\sigma_{n,w}^{(\mathcal M)}} \leq z_\alpha ~\bigg|~ \Delta^{(\mathcal M)}_{w,Z} > 0 \right) \nonumber \\
 &=& \lim_{n\to\infty} \mathbb P\left(\frac{\sqrt{n}\left(\Delta^{(\mathcal M)}_{n,w}-\Delta^{(\mathcal M)}_{w,Z}\right) }{\sigma_{n,w}^{{(\mathcal M)}}} \leq \frac{\sqrt{n} \left(\Delta-\Delta^{(\mathcal M)}_{w,Z}\right) }{\sigma_{n,w}^{{(\mathcal M)}}} + z_\alpha  ~\bigg|~ \Delta^{(\mathcal M)}_{w,Z} > 0 \right) \nonumber \\
 &=& \left\{\begin{array}{ll} 0, & \mbox{if } \Delta^{(\mathcal M)}_{w,Z}>\Delta,\\ \alpha, & \mbox{if} \ \Delta^{(\mathcal M)}_{w,Z}=\Delta, \\ 1, & \mbox{if} \ \Delta^{(\mathcal M)}_{w,Z}<\Delta.
\end{array}\right.
\end{eqnarray}
In  addition, for each law of $Z$ with $\Delta^{(\mathcal M)}_{w,Z}=0$, and by \eqref{asy} and \eqref{varn}, we have
\begin{eqnarray} \label{reject0}
 & & \lim_{n\to\infty} \mathbb P \left( {\rm{Reject}} \ {\cal {H}}_0^{(\mathcal M)} ~\big|~ \Delta^{(\mathcal M)}_{w,Z}=0 \right) \nonumber \\
 &=& \lim_{n\to\infty} \mathbb P\left(\sqrt{n} \Delta^{(\mathcal M)}_{n,w} \leq \sqrt{n} \Delta + \sigma_{n,w}^{{(\mathcal M)}} z_\alpha ~\bigg|~ \Delta^{(\mathcal M)}_{w,Z}=0 \right) \nonumber \\
 &=& 1.
\end{eqnarray}

\begin{remark}\label{remark1}
Note that the first two cases in \eqref{reject} correspond to null hypotheses, while the last case in \eqref{reject} and the case in \eqref{reject0} correspond to alternatives.
Therefore, if \eqref{CLT}--\eqref{varn} hold true, we have
\[
\lim_{n\to\infty} \mathbb P \left( {\rm{Reject}} \ {\cal {H}}_0^{(\mathcal M)} ~\big|~ {\cal {H}}_0^{(\mathcal M)} \right) \leq \alpha, \ \
\lim_{n\to\infty} \mathbb P \left( {\rm{Reject}} \ {\cal {H}}_0^{(\mathcal M)} ~\big|~ {\cal {H}}_1^{(\mathcal M)} \right) = 1,\]
meaning that under the standing conditions, the proposed test defined by the critical region \eqref{CR} is an asymptotic size--$\alpha$ test which is consistent against all fixed alternatives.
\end{remark}

The general theory and specific conditions under which \eqref{CLT}--\eqref{varn} hold true have been presented by using a Hilbert space approach in \cite{Baringhaus2017} and \cite{Henze2020}, for distributional homogeneity and symmetry--equivalence testing, respectively.

Our contribution herein is to introduce a calculus by means of which the test defined in (\ref{CR}) is made feasible.
More precisely by utilizing
the results in \cite{Chen2019} and the asymptotic behavior of $\Delta^{(\mathcal M)}_{n,w}$ obtained within a $U$--statistic framework, we determine specific instances of the test which may be carried out in a computationally friendly way, which also facilitates Monte Carlo approximation and the study of small--sample properties of the resulting tests.

\section{Specification of test statistics} \label{sec3}

For computational convenience as well as for asymptotics we will impose the following assumption on the weight functions.
The generic notation $d$ for dimension is used below, with $d$ varying depending on context.
\begin{itemize}
\item [\rm{(A)}] The weight function  $w(t)$ is the density function of a symmetric around zero distribution on $\mathbb R^d$,
i.e. $w(t) = w(-t), \ t\in\mathbb R^d$, and $\int_{\mathbb R^d} w(t) {\rm{d}}t = 1$.
\end{itemize}
More specifically as weight function we fix the density of a spherically symmetric distribution. The CF of this density will be  denoted by $C_w(\cdot)$, and will be specified further down the paper.

By taking into account the characterization in \cite{Chen2019} we have that, under assumption (A), the population measures in \eqref{CFH}--\eqref{CFI} can be rendered as
\begin{eqnarray} \label{DeltaH}
  \Delta^{(\mathcal H)}_{w,Z} &=& \mathbb E\left[C_w(X-X_1)+C_w(Y-Y_1)-2 C_w(X-Y)\right],
\end{eqnarray}
\begin{eqnarray} \label{DeltaS}
  \Delta^{(\mathcal S)}_{w,Z} &=& \frac{1}{2}\mathbb E\left [C_w(X-X_1)-C_w(X+X_1)\right],
\end{eqnarray}
\begin{eqnarray} \label{DeltaI}
  \Delta^{(\mathcal I)}_{w,Z} &=& \mathbb E\left[C_{w_p}(X-X_1)C_{w_q}(Y-Y_1)\right] + \mathbb E\left[C_{w_p}(X-X_1)\right]\mathbb E\left[C_{w_q}(Y-Y_1)\right] \\ \nonumber
  & & - \ 2\mathbb E\left[C_{w_p}(X-X_1)C_{w_q}(Y-Y_2)\right],
\end{eqnarray}
where $Z_1$ and $Z_2$ denote independent copies of the random vector $Z$.
Thus, it is natural to estimate the population measure $\Delta^{(\mathcal M)}_{w,Z}$ by means of
suitable $U$--statistics, whose corresponding kernels are determined by $C_w(\cdot)$; see, e.g., \cite{Hoeffding1948, Serfling1980}.

To begin with the symmetry statistic, suppose that $\{X_1,\ldots,X_n\}$ are independent copies of $X$, the empirical counterpart of $\Delta^{(\mathcal S)}_{w,Z}$ may be obtained as
\begin{eqnarray} \label{testSU}
 \Delta^{(\mathcal S)}_{n,w} = \frac{1}{2n(n-1)} \sum_{1\leq i\neq j\leq n} \left\{ C_w(X_i-X_j)-C_w(X_i+X_j) \right\},
\end{eqnarray}
i.e. as a $U$--statistic with symmetric kernel $\psi^{\mathcal S}(x,x_1)=\{C_w(x-x_1)-C_w(x+x_1)\}/2$.

Likewise suppose that in addition to  $\{X_1,\ldots,X_n\}$,  and independently, we have $\{Y_1,\ldots,Y_n\}$ as independent copies of $Y$. Then the empirical counterpart of $\Delta^{(\mathcal H)}_{w,Z}$ may be formulated as
\begin{eqnarray} \label{testHU}
 \Delta^{(\mathcal H)}_{n,w} = \frac{1}{n(n-1)} \sum_{1\leq i\neq j\leq n} \left\{ C_w(X_i-X_j)+C_w(Y_i-Y_j)-2C_w(X_i-Y_j) \right\},
\end{eqnarray}
which is a $U$--statistic with symmetric kernel $\psi^{\mathcal H}(z,z_1)=C_w(x-x_1)+C_w(y-y_1)-C_w(x-y_1)-C_w(x_1-y)$. Here, $z=(x^\top,y^\top)^\top$.

In turn if $\{(X_1,Y_1),\ldots,$ $(X_n,Y_n)\}$ are independent copies of $(X,Y)$, the empirical counterpart of $\Delta^{(\mathcal I)}_{w,Z}$ is given by
\begin{eqnarray}  \label{testIU}
\Delta^{(\mathcal I)}_{n,w} &=& \frac{1}{n(n-1)} \sum_{1\leq i\neq j\leq n} C_{w_p}(X_i-X_j)C_{w_q}(Y_i-Y_j) \\ \nonumber
               & & + \ \left\{\frac{1}{n(n-1)} \sum_{1\leq i\neq j\leq n} C_{w_p}(X_i-X_j)\right\} \left\{ \frac{1}{n(n-1)} \sum_{1\leq i\neq j\leq n} C_{w_q}(Y_i-Y_j) \right\} \\ \nonumber
               & & - \ \frac{2}{n(n-1)(n-2)} \sum_{1\leq i\neq j\neq k\leq n} C_{w_p}(X_i-X_j)C_{w_q}(Y_i-Y_k) \\ \nonumber
               &:=& U_{n,1} + U_{n,2}U_{n,3} -2U_{n,4}.
\end{eqnarray}
Denote $S_1=\mathbb E\left[C_{w_p}(X-X_1)C_{w_q}(Y-Y_1)\right]$, $S_2=\mathbb E\left[C_{w_p}(X-X_1)\right]$, $S_3=\mathbb E\left[C_{w_q}(Y-Y_1)\right]$, and $S_4=\mathbb E\left[C_{w_p}(X-X_1)C_{w_q}(Y-Y_2)\right]$.
Thus each $U_{n,i}$ ($i=1,\ldots,4$) is a $U$--statistic of degree $m_i$ for estimation of $S_i$ based on the symmetric kernel $\psi^{\mathcal I,i}$, where $m_1=m_2=m_3=2$, $m_4=3$,
$\psi^{\mathcal I,1}(z,z_1) = C_{w_p}(x-x_1)C_{w_q}(y-y_1),$
$\psi^{\mathcal I,2}(z,z_1) = C_{w_p}(x-x_1),$
$\psi^{\mathcal I,3}(z,z_1) = C_{w_q}(y-y_1),$
and
\begin{eqnarray*}
\psi^{\mathcal I,4}(z,z_1,z_2) &=& \frac{1}{6} \Big\{ C_{w_p}(x-x_1)C_{w_q}(y-y_2) + C_{w_p}(x-x_2)C_{w_q}(y-y_1) \\ \nonumber
 & & + \ C_{w_p}(x-x_1)C_{w_q}(y_1-y_2) + C_{w_p}(x_1-x_2)C_{w_q}(y-y_1) \\ \nonumber
 & & + \ C_{w_p}(x-x_2)C_{w_q}(y_1-y_2) + C_{w_p}(x_1-x_2)C_{w_q}(y-y_2) \Big\}.
\end{eqnarray*}
Here $\psi^{\mathcal I,4}$ is a symmetrized version of the asymmetric kernel $C_{w_p}(x-x_1)C_{w_q}(y-y_2)$ of $S_4$.
Consequently, $\Delta^{(\mathcal I)}_{n,w}$ is a function of the $U$--statistics $U_{n,1},\ldots,U_{n,4}$.
Specifically if we let $h(u)=u_1+u_2u_3-2u_4$ be a known real--valued function of a four--dimensional vector $u$, then $\Delta^{(\mathcal I)}_{w,Z} =h(S)$ and $\Delta^{(\mathcal I)}_{n,w} = h(U_n)$, where $S=(S_1,\ldots,S_4)$ and $U_n=(U_{n,1},\ldots,U_{n,4})$.

According to the above discussion, the asymptotic behavior of $\Delta^{(\mathcal S)}_{n,w}$ and $\Delta^{(\mathcal H)}_{n,w}$ can be deduced from the general theory of $U$--statistics (see, e.g., \cite{Hoeffding1948, Serfling1980}). Likewise, the asymptotic behavior of $\Delta^{(\mathcal I)}_{n,w}$
can be obtained from the asymptotic properties of functions of several $U$--statistics (see e.g., \cite{Hoeffding1948, Lee1990}).
We now formulate the asymptotic properties of $\Delta^{(\mathcal M)}_{n,w}$ in \eqref{testSU}--\eqref{testIU} in the following theorem uniformly. Again the composition of the random vector $Z$ below varies depending on the context. Also recall that in the context of homogeneity--equivalence test, $Z=\{X,Y\}$, with $X$ and $Y$ assumed independent, but we will reconsider this assumption in Remark \ref{rem1}.
\begin{theorem} \label{thm}
Let $\{Z_1,\ldots,Z_n\}$ be independent copies of $Z$. Then under assumption (A), and as $n\rightarrow \infty$, we have the following results:
\begin{enumerate}
  \item[(i)] If $\Delta^{(\mathcal M)}_{w,Z} > 0$, then
  \begin{eqnarray*} \label{CLTM}
    \sqrt{n} \left(\Delta^{(\mathcal M)}_{n,w}-\Delta^{(\mathcal M)}_{w,Z} \right) \stackrel{{\cal D}}{\longrightarrow} \mathcal N\left(0,\sigma_w^{2(\mathcal M)}\right),
  \end{eqnarray*}
  with the limit variance $\sigma_w^{2(\mathcal M)}$, respectively  given by,
  \[ \sigma_w^{2(\mathcal S)} = 4 \left\{ \mathbb E\left[ \psi^{\mathcal S}(X,X_1)\psi^{\mathcal S}(X,X_2) \right] - \mathbb E^2 \left[ \psi^{\mathcal S}(X,X_1) \right] \right\}, \]
  \[ \sigma_w^{2(\mathcal H)} = 4 \left\{ \mathbb E\left[ \psi^{\mathcal H}(Z,Z_1)\psi^{\mathcal H}(Z,Z_2) \right] - \mathbb E^2 \left[ \psi^{\mathcal H}(Z,Z_1) \right] \right\}, \]
 \[ \sigma_w^{2(\mathcal I)} = \sum_{i,j=1}^4 m_i m_j h'_i(S) h'_j(S) \zeta_1(i,j), \]
 where $h'_i(S):=\frac{\partial h}{\partial u_i} |_{u=S}$ denote the first-order partial derivative of $h$, and $\zeta_1(i,j)=\zeta_1(j,i)$ are defined in the Appendix.
 \item[(ii)] If $\Delta^{(\mathcal M)}_{w,Z} = 0$, then
  \begin{eqnarray*} \label{asyM}
    \sqrt{n} \Delta^{(\mathcal M)}_{n,w} \stackrel{\mathbb  P}{\longrightarrow} 0
  \end{eqnarray*}
\end{enumerate}
\end{theorem}

Theorem \ref{thm} presents the specific conditions under which \eqref{CLT} and \eqref{asy} hold true. Now we turn to study  consistent estimation of the limit variance $\sigma_w^{2(\mathcal M)}$.
We first define estimators of $\sigma_w^{2(\mathcal S)}$ and $\sigma_w^{2(\mathcal H)}$, which are subsequently shown to be consistent. The proof is based on the strong law of large numbers (SLLN) for $U$--statistics (\cite[see][Theorem 5.4A]{Serfling1980}) and Slutsky's theorem.

\begin{theorem} \label{thm2}
Let $\{X_1,\ldots,X_n\}$ and $\{Z_1,\ldots,Z_n\}$ be independent copies of $X$ and $Z$ respectively, and define
\begin{eqnarray} \label{varnS}
  \sigma_{n,w}^{2(\mathcal S)} &=& \frac{4}{n(n-1)(n-2)} \sum_{1\leq i\neq j\neq k\leq n} \psi^{\mathcal S}(X_i,X_j)\psi^{\mathcal S}(X_i,X_k) \\ \nonumber
  & & - \ 4 \left\{ \frac{1}{n(n-1)} \sum_{1\leq i\neq j\leq n} \psi^{\mathcal S}(X_i,X_j) \right\}^2,
\end{eqnarray}
and
\begin{eqnarray} \label{varnH}
  \sigma_{n,w}^{2(\mathcal H)} &=& \frac{4}{n(n-1)(n-2)} \sum_{1\leq i\neq j\neq k\leq n} \psi^{\mathcal H}(Z_i,Z_j)\psi^{\mathcal H}(Z_i,Z_k) \\ \nonumber
  & & - \ 4 \left\{ \frac{1}{n(n-1)} \sum_{1\leq i\neq j\leq n} \psi^{\mathcal H}(Z_i,Z_j) \right\}^2.
\end{eqnarray}
Then as $n\rightarrow \infty$, we have
\begin{eqnarray*}
   \sigma_{n,w}^{2(\mathcal S)} \stackrel{\mathbb  P}{\longrightarrow} \sigma_w^{2(\mathcal S)}, \ \ \ \ \ \ \sigma_{n,w}^{2(\mathcal H)} \stackrel{\mathbb  P}{\longrightarrow} \sigma_w^{2(\mathcal H)},
\end{eqnarray*}
where $\sigma_w^{2(\mathcal S)}$ and $\sigma_w^{2(\mathcal H)}$ are defined in Theorem \ref{thm}.
\end{theorem}


The case of independence however is somewhat different in that in order to obtain an estimator of $\sigma_w^{2(\mathcal I)}$ analogous to \eqref{varnS} and \eqref{varnH}, the number of terms to be estimated in $\sigma_w^{2(\mathcal I)}$ is equal to $4(4+1)/2 = 10$. Therefore, an analogous estimator of $\sigma_w^{2(\mathcal I)}$ is very complicated to simulate.
Instead, we will estimate the variance of $\Delta^{(\mathcal I)}_{n,w} = h(U_n)$ by jackknife, which is a widely applicable statistical tool used for estimating variance; see \cite{Efron1982} for a full account of the jackknife.

\begin{theorem} \label{thm3}
Let $\{(X_1^\top,Y_1^\top)^\top,\ldots,(X_n^\top,Y_n^\top)^\top\}$ be independent copies of $(X^\top,Y^\top)^\top$, and define
\begin{eqnarray} \label{varnI}
  \sigma_{n,w}^{2(\mathcal I)} &=& (n-1)\sum_{i=1}^n \Big( h(U_{n-1}(-i)) - \bar{h}(U_{n-1}) \Big)^2,
\end{eqnarray}
where $U_n=(U_{n,1},\ldots,U_{n,4})$, $U_{n,1},\ldots,U_{n,4}$ are defined in \eqref{testIU}, $h(u)=u_1+u_2u_3-2u_4$ be a known real--valued function of a four--dimensional vector $u$, $h(U_{n-1}(-i))$, $i=1,\ldots,n$, denote the leave--one--out statistics that are computed from the sample of $n-1$ observations formed by deleting the $i^{\rm{th}}$ observation,
and $\bar{h}(U_{n-1}) = n^{-1}\sum_{i=1}^n h(U_{n-1}(-i))$.
Then as $n\rightarrow \infty$, we have
\begin{eqnarray*}
   \sigma_{n,w}^{2(\mathcal I)} \stackrel{\mathbb  P}{\longrightarrow} \sigma_w^{2(\mathcal I)},
\end{eqnarray*}
where $\sigma_w^{2(\mathcal I)}$ is defined in Theorem \ref{thm}.
\end{theorem}

\begin{remark}\label{remark2}
Theorems \ref{thm}--\ref{thm3} present the specific conditions under which \eqref{CLT}--\eqref{varn} hold true for the three model--equivalence tests. As a consequence, see also the  discussion in Remark \ref{remark1}, the tests defined by the critical region \eqref{CR} are asymptotic size--$\alpha$ tests and consistent against all fixed alternatives.
\end{remark}

\begin{remark} \label{rem1}
It should be pointed out that even without the assumption of independence of $X$ and $Y$, it holds
\[
 \Delta^{(\mathcal H)}_{w,Z} = \mathbb E[C_w(X-X_1)+C_w(Y-Y_1)-C_w(X-Y_1)-C_w(Y-X_1)],
\]
in the case of homogeneity testing. This has the same symmetric kernel $\psi^{\mathcal H}(z,z_1)$ with that in \eqref{testHU}, and opens up the possibility of extending the method to a more general two--sample problem with dependent observations. Here however we remain strictly within the classical two--sample problem under which $X$ and $Y$ are independent. For more information on the case of dependent observations we refer to \cite{Quessy2012, Hlavka2020}.
\end{remark}

We close this section by specifying the spherical densities which will be employed in our simulation study. In this connection note that in principle  any spherical density (or in fact any centrally symmetric density) may be used as weight function. However  there exist some choices that are convenient from the computational point of view. Specifically we will employ as weight function the density of the spherical stable distribution and the density of the generalized spherical Laplace distribution. The CFs of these densities will be denoted by $C_\gamma(\cdot)$, where $C_\gamma(t)=\exp(-\|t\|^\gamma), \ \gamma\in(0,2]$,  corresponds to the stable density, while $C_\gamma(t)=(1+\|t\|^2)^{-\gamma}, \ \gamma>0$, corresponds to the Laplace density; see \cite{Nolan2013} and \cite{Kozubowski2013}, respectively. Here $\gamma$ may be termed  ``characteristic exponent". With these densities as weight functions, formulae for the sample measures $\Delta^{(\mathcal M)}_{n,Z}$ in \eqref{testSU}--\eqref{testIU} as well as the estimated limit variance $\sigma_{n,w}^{2(\mathcal M)}$ in
\eqref{varnS}--\eqref{varnI}, readily result by simply replacing $C_w(\cdot)$ by $C_\gamma(\cdot)$ in the corresponding equations.
(Recall though that in the case of independence testing and due to the complexity of the resulting formulae, we employ the jackknife in order to estimate the limit variance).
\begin{remark}
The weight function $w(t)$ figuring in our population measures in \eqref{CFH}--\eqref{CFI} as well as in our tests statistics  may in principle take arbitrary functional forms. Trivial conditions are that $w(t)$ should be non-negative and symmetric around zero. A further requirement, already mentioned, is computational convenience, meaning that it should render kernels $C_w(\cdot)$ that are in a closed--form and free of numerical integration. In this connection several authors have discussed weight functions that lead to so--called universal kernels, i.e. kernels that lead to test consistency; see \cite{Gretton2012} for related work in the setting  of reproducing kernel Hilbert spaces.  In fact \cite{Micchelli2006} already mention the normal and (generalized) Laplace as universal kernels. The kernels corresponding to the energy--distance statistics are also universal; see for instance \cite{Sejdinovic2013}.
The advantage of density--based weight functions adopted herein is that energy statistics, by  using a non--integrable weight function rather than a density,  impose certain moment conditions on the underlying random variable(s) that restricts application to light--tailed distributions.
\end{remark}

\section{Numerical studies} \label{sec4}

In this section, a series of simulation experiments  are carried out in order to investigate the finite--sample performance of the proposed tests.
As mentioned at the end of Section \ref{sec3}, we employ the stable density and the Laplace density as the weight function, and the corresponding tests are denoted as $T^{(\mathcal M)}_{n,S}$ and $T^{(\mathcal M)}_{n,L}$ respectively.
As for the values of the characteristic exponent, we choose $\gamma \in \{0.5,1.0,1.5,2.0\}$ for the stable density and $\gamma \in \{0.1,0.25,1.0,4.0\}$ for the Laplace density.
In the following, empirical rejection rates according to the critical regions \eqref{CR} are calculated respectively at $5\%$ significance level using Monte Carlo $2000$ trials.
Further, the dimensions of $X$ (and $Y$) considered are $p$ ($=q$) $\in \{2,4,6\}$, and the sample sizes are $n \in \{50, 100,200,300\}$.

We also wish to compare the new procedures with competitors. In this connection for testing symmetry we refer to
\cite{Meintanis2012}, while for testing homogeneity to \cite{Alba2008, Quessy2012, Ghosh2016, Alba2017}, and for testing independence to \cite{Genest2004, Zhang2018}, all in the standard setting of testing exact hypotheses. In the existing literature however we did not find corresponding  methods for testing in the context of multivariate equivalence hypotheses. Nevertheless the energy statistics of Sz\'ekely and Rizzo \cite{Szekely2013} have been adapted herein to the context of model--equivalence testing for comparison purposes. For these tests we write $T^{(\mathcal M)}_{n,E}$. In this connection, the energy--based homogeneity population measure defined in \eqref{szekely} may formally be obtained by replacing $C_w(\cdot)$ with $-\|\cdot\|^\gamma$ in \eqref{DeltaH}, and consequently the test $T^{(\mathcal H)}_{n,E}$ can be implemented by simply replacing $C_w(\cdot)$ with $-\|\cdot\|^\gamma$ in \eqref{testHU}.
With analogous replacements one can obtain from \eqref{DeltaS} and \eqref{DeltaI} corresponding energy population measures for symmetry and independence, respectively.
Furthermore, within a $U$--statistic framework, the asymptotic properties of the energy--based test statistics analogous to \eqref{CLT}--\eqref{varn} can be obtained when the moment conditions $\mathbb E [ \|X\|^{2\gamma} ] < \infty$ and $\mathbb E [\|Y\|^{2\gamma} ] < \infty$ hold true.
However, it is worth noticing that 
the proposed test statistics in this paper require no moment conditions.
In what follows we compare our procedures with their energy counterparts $T^{(\mathcal M)}_{n,E}$, for $\gamma \in \{0.5,1.0,1.5,2.0\}$ (note that although the value $\gamma=2.0$ is excluded from the energy--based characterizations we nevertheless include it in our simulations).

\textbf{Example 1.} Symmetry--equivalence test:
We consider two scenarios in this example, with $X$ generated from the multivariate skew-normal distribution $\mathcal{SN}_p(\xi, \Omega,\delta)$ in Example 1(a) and from the multivariate skew-Cauchy distribution $\mathcal{SC}_p(\xi, \Omega,\delta)$ in Example 1(b), respectively.
Here, $\xi$ is a $p$--dimensional vector representing the location parameter, $\Omega$ is a symmetric positive definite $p \times p$ scatter matrix, and $\delta$ is a $p$--dimensional vector which regulates the skewness of the distributions; see the monograph \cite{Azzalini2014} and the \textsf{R} package \texttt{sn} (\cite{Azzalini2022}) for additional information.
In Example 1(a), we choose $X \sim \mathcal{SN}_p(0_p, I_p,\theta 1_p)$, where $0_p$ and $1_p$ are $p$--dimensional vectors of zeros and ones respectively, $I_p$ denotes the identity matrix, and $\theta \geq 0$ is a scalar.
With this parametrization, the degree of asymmetry can be measured  in terms of $\theta$, in the sense that smaller $\theta$ corresponds to less asymmetry.
Particularly, $\mathcal{SN}_p(0_p, I_p,\theta 1_p)$ reduces to the multivariate standard normal distribution when $\theta=0$.
We choose some fixed value $\theta=\theta_0$ as benchmark, and compute the value of $\Delta^{(\mathcal S)}_{w,X}$ in \eqref{DeltaS}
for $X\sim\mathcal{SN}_p(0_p, I_p,\theta_01_p)$ as a general threshold $\Delta$ in \eqref{CR} for significant departure from symmetry.
Thus distributions $\mathcal{SN}_p(0_p, I_p,\theta 1_p)$ with $\theta \geq \theta_0$ correspond to the null hypotheses and we are in the alternatives when $\theta < \theta_0$.
A random approximation procedure is used to compute $\Delta^{(\mathcal S)}_{w,X}$ for $\mathcal{SN}_p(0_p, I_p,\theta_01_p)$.
Specifically, we generate a large number $B$ of i.i.d. copies  $\{X_1, \ldots, X_B\}$ of $X\sim\mathcal{SN}_p(0_p, I_p,\theta_01_p)$, and calculate $\Delta^{(\mathcal S)}_{n,w}$ in \eqref{testSU} based on this sample as the threshold $\Delta$. 
To ensure the accuracy of the approximation,  $B=5000$ is used, which is a fairly large value; see the closing paragraph of this section for more information on the accuracy of the random approximation method. As benchmark we choose $\mathcal{SN}_p(0_p, I_p,\theta_01_p)$ with $\theta_0=3$, and also consider $\theta \in \{5,4,3,2,1,0\}$.
In Example 1(b), we repeat Example 1(a) under identical conditions except that $X\sim\mathcal{SC}_p(0_p, I_p,\theta1_p)$.

\begin{figure}[b]
\centering
\includegraphics[scale=0.7]{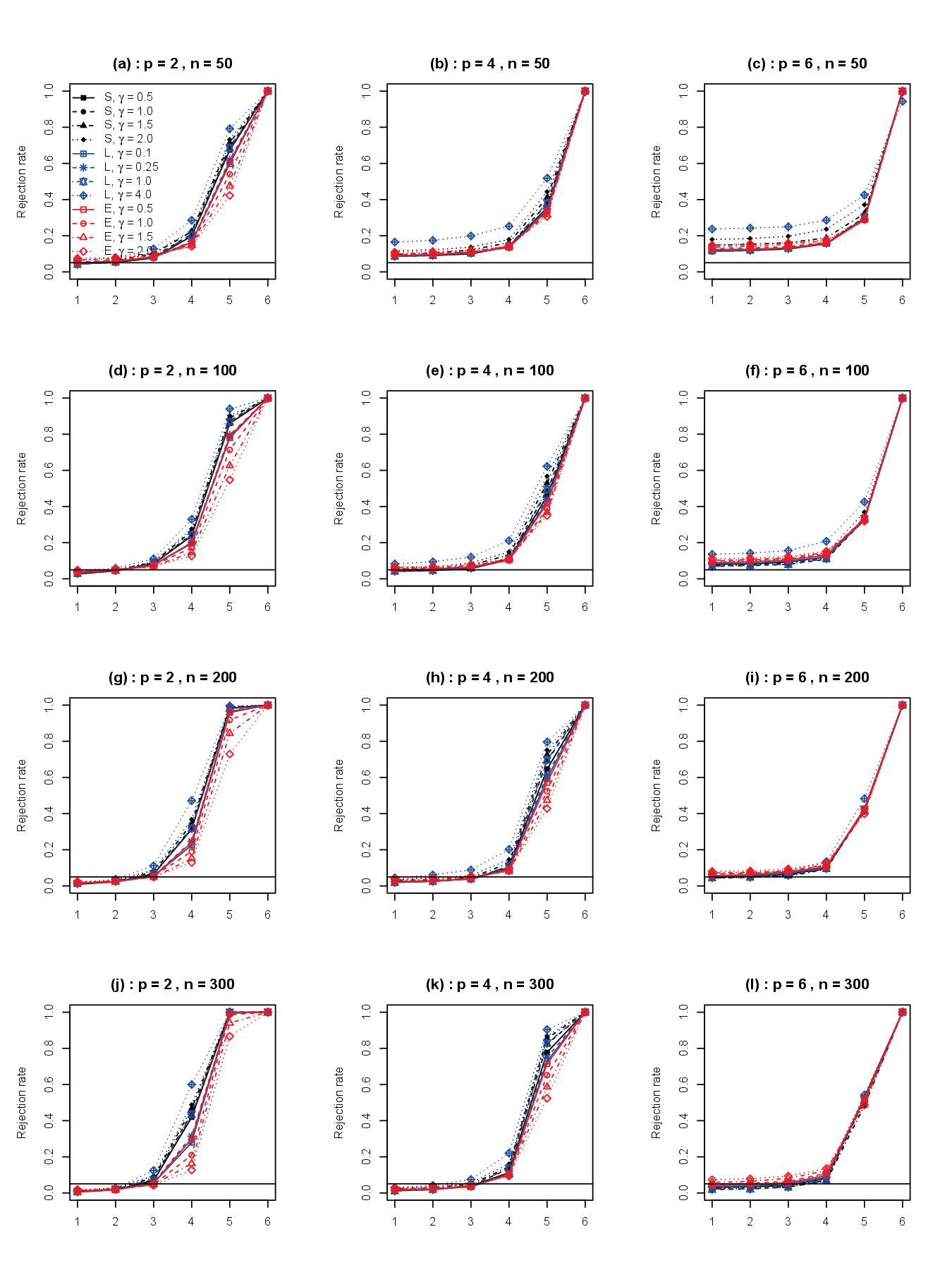}\\
\vspace{-1cm}
\caption{Simulation results for Example 1(a) based on $2000$ trials. The horizontal line corresponds to the $5\%$ significance level. The capital letters S, L, and E in panel (a) represent $T^{(\mathcal S)}_{n,S}$, $T^{(\mathcal S)}_{n,L}$, and $T^{(\mathcal S)}_{n,E}$ respectively.} \label{figS1}
\end{figure}

\begin{figure}[b]
\centering
\includegraphics[scale=0.7]{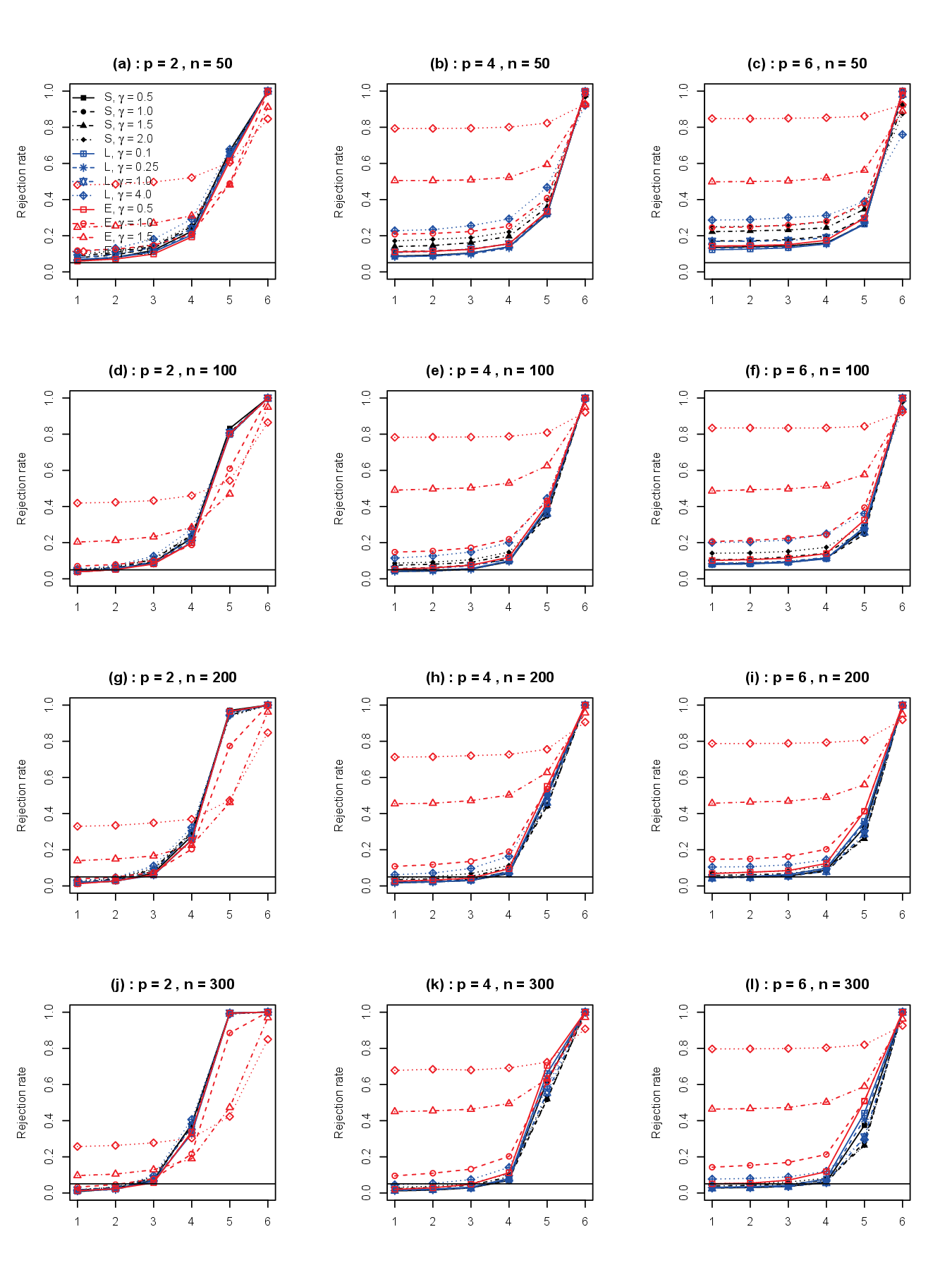}\\
\vspace{-1cm}
\caption{Simulation results for Example 1(b) based on $2000$ trials. The horizontal line corresponds to the $5\%$ significance level. The capital letters S, L, and E in panel (a) represent $T^{(\mathcal S)}_{n,S}$, $T^{(\mathcal S)}_{n,L}$, and $T^{(\mathcal S)}_{n,E}$ respectively.} \label{figS2}
\end{figure}

The empirical rejection rates of the proposed symmetry--equivalence test with the test $T^{(\mathcal S)}_{n,S}$ corresponding to the spherical stable density as weight function are shown in Table \ref{tableS2}.
We also display the comparative results of all three tests $T^{(\mathcal S)}_{n,S}$, $T^{(\mathcal S)}_{n,L}$, and $T^{(\mathcal S)}_{n,E}$ for Examples 1(a) and 1(b) in Figures \ref{figS1} and \ref{figS2}, respectively.

\begin{center}
  Table \ref{tableS2} and Figures \ref{figS1}, \ref{figS2} about here
\end{center}

In general, the performance of the three tests $T^{(\mathcal S)}_{n,S}$, $T^{(\mathcal S)}_{n,L}$, and $T^{(\mathcal S)}_{n,E}$ are comparable in Example 1(a).
However, in Example 1(b), where $X\sim\mathcal{SC}_p(0_p, I_p,\theta1_p)$ and the expectation of $X$ does not exist, the energy--based test $T^{(\mathcal S)}_{n,E}$ with $\gamma=1.5$ or $\gamma=2.0$ can not control the Type--I error rates, even when $p=2, n=300$.
We may  attribute this behavior to the fact that in this case the moment condition corresponding to the energy statistic is not satisfied.
On the other hand, it is clear that the Type--I error rates of the proposed tests $T^{(\mathcal S)}_{n,S}$ and $T^{(\mathcal S)}_{n,L}$ can be controlled well for lower dimensions ($p=2, 4$) with all sample sizes and a higher dimension ($p=6$) with a larger sample size ($n=300$).
Furthermore, it can be seen that, for all cases, the Type--I error rates in the interior of the null hypothesis (i.e., $\theta = 5, 4$) are lower than those at the boundary of the null hypothesis (i.e., $\theta_0=3$), which is line with the limit result in \eqref{reject}.
As for the empirical powers, it may be observed that with the increase of the sample size and/or distance away from the null hypothesis, the empirical powers increase with a few exceptions, which again is expected in view of our consistency results.
Finally, the value of $\gamma$ is seen to affect the empirical powers of all three tests, but only slightly. 

\textbf{Example 2.} Homogeneity--equivalence test:
Two scenarios are considered in this example. In Example 2(a), we consider an independent pair of random vectors such that $X\sim \mathcal{N}_p(0,1)$ and $Y\sim \mathcal{N}_p(\mu,1)$, where $\mathcal{N}_p(\mu, \sigma^2)$ denotes a $p$--dimensional Gaussian random vector whose coordinates are independent and identically distributed as univariate normal distributions with mean $\mu$ and variance equal to $\sigma^2$.
We also choose some fixed value of $\mu=\mu_0$ as a benchmark, and set the corresponding value of $\Delta^{(\mathcal H)}_{w,Z}$ in \eqref{DeltaH}
as a general threshold $\Delta$ in \eqref{CR} for significant departure from distributional homogeneity.
Thus $\mu \geq \mu_0$ and $\mu<\mu_0$ correspond to the null hypotheses and alternatives, respectively.
Also, the benchmark threshold $\Delta$ is computed according to a random approximation procedure.
Furthermore, the benchmark and non--benchmark values considered are $\mu_0=2.0$ and $\mu \in \{2.2,2.1,2.0,1.9,1.8,1.7\}$.
In Example 2(b), we repeat Example 2(a) under identical conditions except that the coordinates of $X$ and $Y$ are independently generated from $G(5,1)$ and $G(5,\mu)$, where $G(a,s)$ denotes the univariate Gamma distribution with shape parameter $a$ and scale parameter $s$.

\begin{figure}[b]
\centering
\includegraphics[scale=0.7]{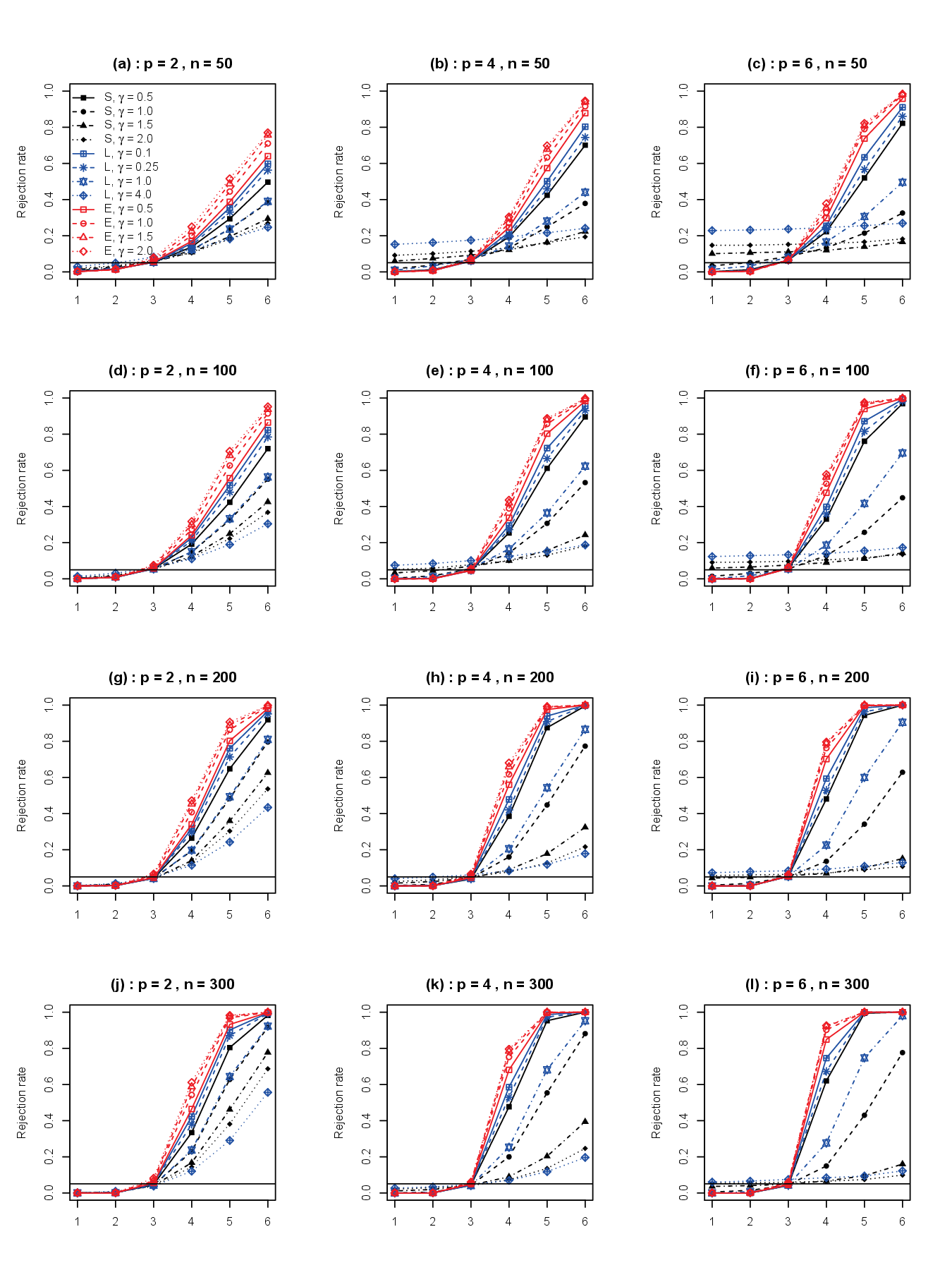}\\
\vspace{-1cm}
\caption{Simulation results for Example 2(a) based on $2000$ trials. The horizontal line corresponds to the $5\%$ significance level. The capital letters S, L, and E in panel (a) represent $T^{(\mathcal H)}_{n,S}$, $T^{(\mathcal H)}_{n,L}$, and $T^{(\mathcal H)}_{n,E}$ respectively.} \label{figH1}
\end{figure}

\begin{figure}[b]
\centering
\includegraphics[scale=0.7]{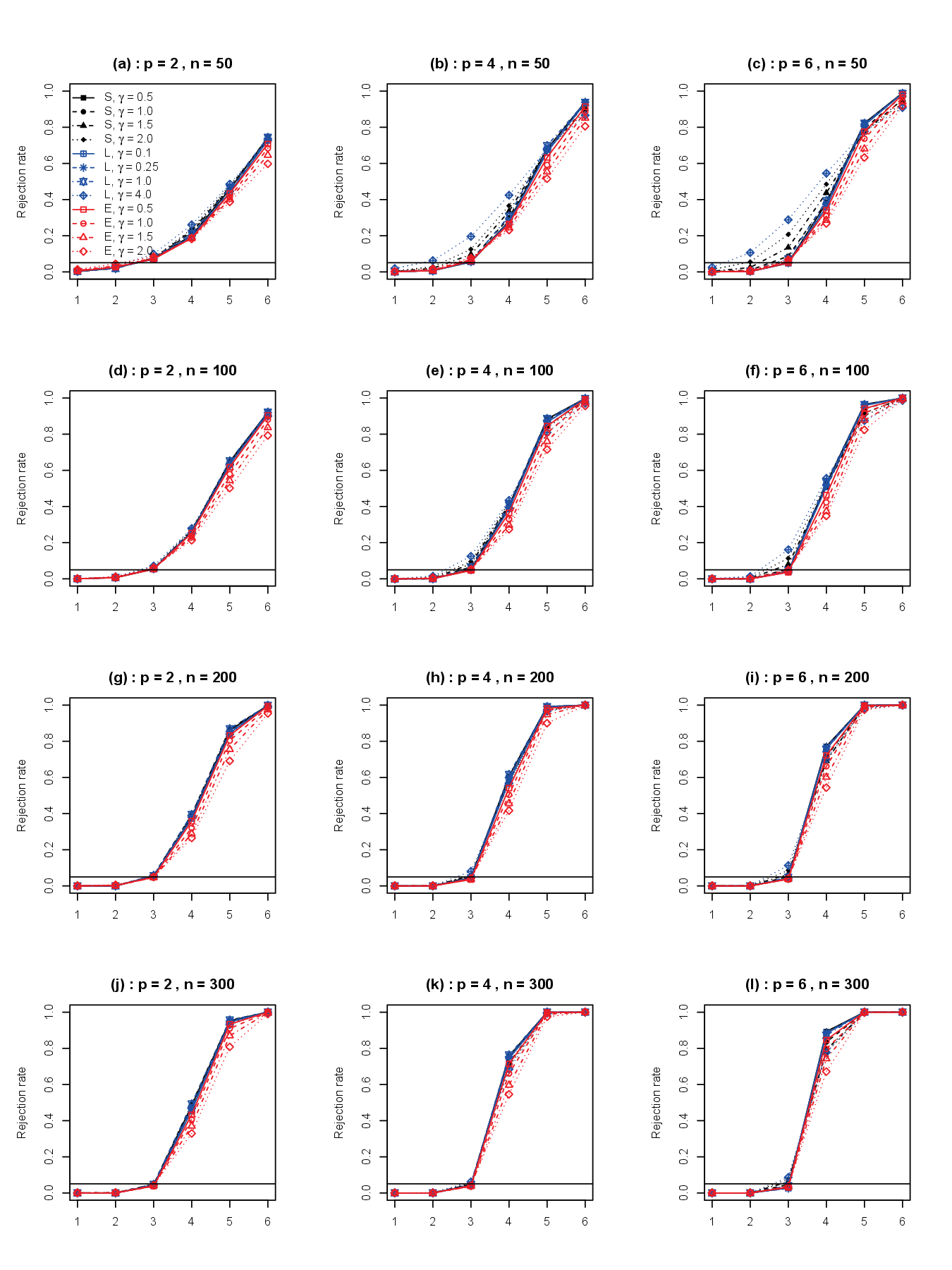}\\
\vspace{-1cm}
\caption{Simulation results for Example 2(b) based on $2000$ trials. The horizontal line corresponds to the $5\%$ significance level. The capital letters S, L, and E in panel (a) represent $T^{(\mathcal H)}_{n,S}$, $T^{(\mathcal H)}_{n,L}$, and $T^{(\mathcal H)}_{n,E}$ respectively.} \label{figH2}
\end{figure}

The resulting empirical rejection rates for $T^{(\mathcal H)}_{n,S}$ are reported in Table \ref{tableH2} while corresponding comparative results of all three tests $T^{(\mathcal H)}_{n,S}$, $T^{(\mathcal H)}_{n,L}$, and $T^{(\mathcal H)}_{n,E}$ are displayed in Figures \ref{figH1} and \ref{figH2}.
It may be seen clearly that the Type--I error rates are generally controlled with a few exceptions appearing when $p$ is higher and $n$ is smaller, and that empirical rejection rates under alternatives are in line with the fact that the tests are consistent.
In addition, the energy--based test $T^{(\mathcal H)}_{n,E}$ is more powerful in Example 2(a), whereas in Example 2(b), the opposite is true.
Furthermore in Example 2(a), and as it may be seen from Figure \ref{figH1}, the value of $\gamma$  has a certain effect on the empirical powers of the proposed tests $T^{(\mathcal H)}_{n,S}$ and $T^{(\mathcal H)}_{n,L}$.  
We postpone further discussion on $\gamma$ till the end of this section.
\begin{center}
  Table \ref{tableH2} and Figures \ref{figH1}, \ref{figH2} about here
\end{center}

\textbf{Example 3.} Independence--equivalence test:
We consider two scenarios in this example, with $Z \sim \mathcal{MN}(0_{p+q}, \Sigma_\rho)$ in Example 3(a) and $Z \sim \mathcal{MT}(0_{p+q}, \Sigma_\rho, 5)$ in Example 3(b), respectively.
Here, $Z=(X^\top,Y^\top)^\top$, $\mathcal{MN}(\mu, \Sigma)$ denotes the multivariate normal distribution with mean vector $\mu$ and covariance matrix $\Sigma$, and $\mathcal{MT}(\delta, \Sigma, \nu)$ denotes the multivariate $t$ distribution (\cite{Kotz2004}) with non-centrality parameter $\delta$, scale matrix $\Sigma$ and degrees of freedom $\nu$. Furthermore,
$$ \Sigma_\rho = \left(
              \begin{array}{cc}
                I_p & \Sigma_0 \\
                \Sigma_0^\top & I_q \\
              \end{array}
            \right)
$$
is a $(p+q) \times (p+q)$ matrix, and $\Sigma_0$ is a $p \times q$ matrix with the $(i,j)^{\rm{th}}$ element equals to $\rho\times \delta_{ij}$ for all $i=1,\ldots,p$ and $j=1,\ldots,q$, where $\delta_{ij}=1$ if $i=j$ and $\delta_{ij}=0$, otherwise.
Thus $\rho$ quantifies the dependence between $X$ and $Y$.
We again choose some fixed value of $\rho=\rho_0$ as a benchmark, and set the corresponding value of $\Delta^{(\mathcal I)}_{w,Z}$ in \eqref{DeltaI} as $\Delta$, with $\Delta$ being approximated by means of the random approximation procedure. 

\begin{figure}[b]
\centering
\includegraphics[scale=0.7]{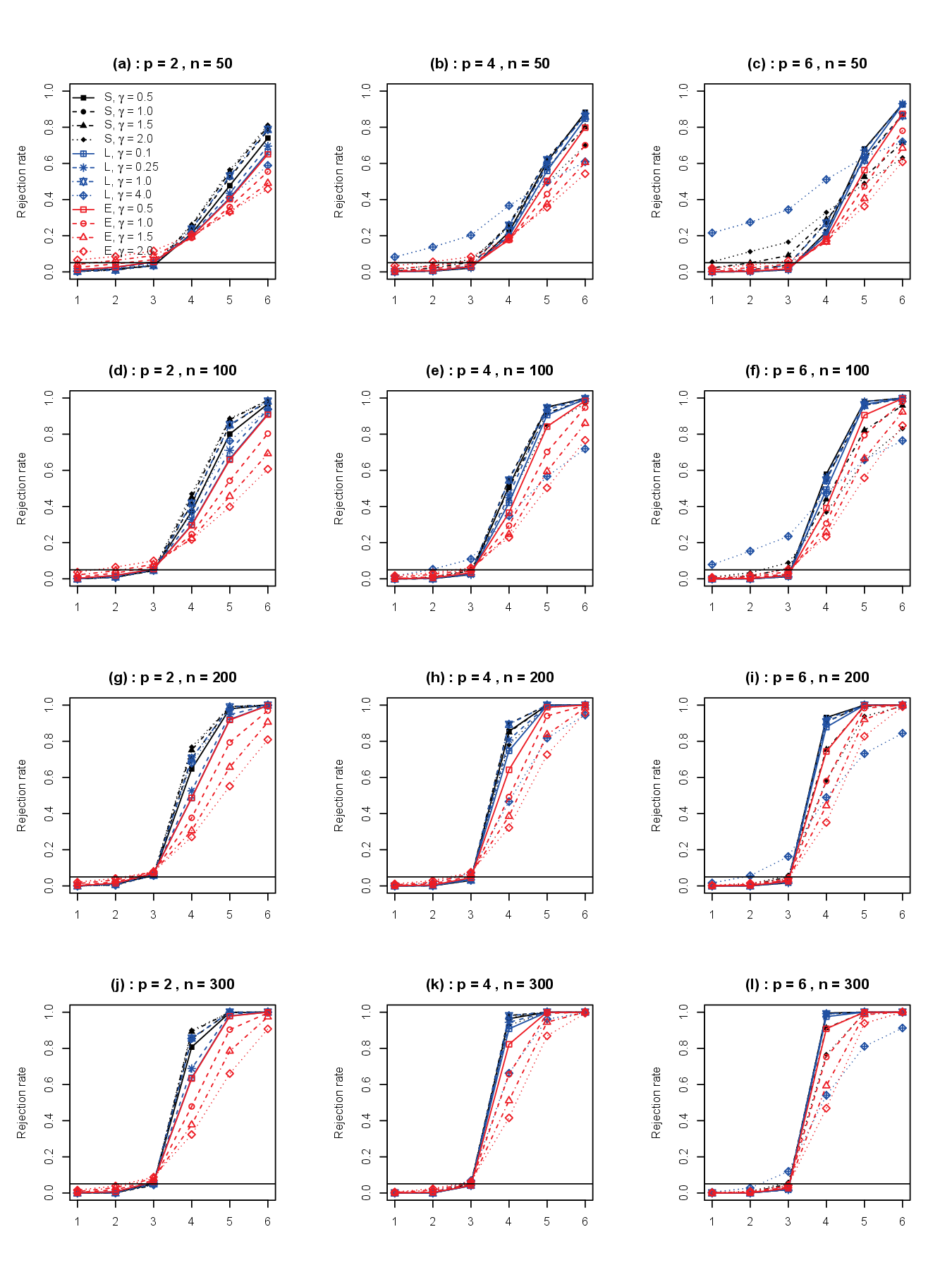}\\
\vspace{-1cm}
\caption{Simulation results for Example 3(a) based on $2000$ trials. The horizontal line corresponds to the $5\%$ significance level. The capital letters S, L, and E in panel (a) represent $T^{(\mathcal I)}_{n,S}$, $T^{(\mathcal I)}_{n,L}$, and $T^{(\mathcal I)}_{n,E}$ respectively.} \label{figI1}
\end{figure}

\begin{figure}[b]
\centering
\includegraphics[scale=0.7]{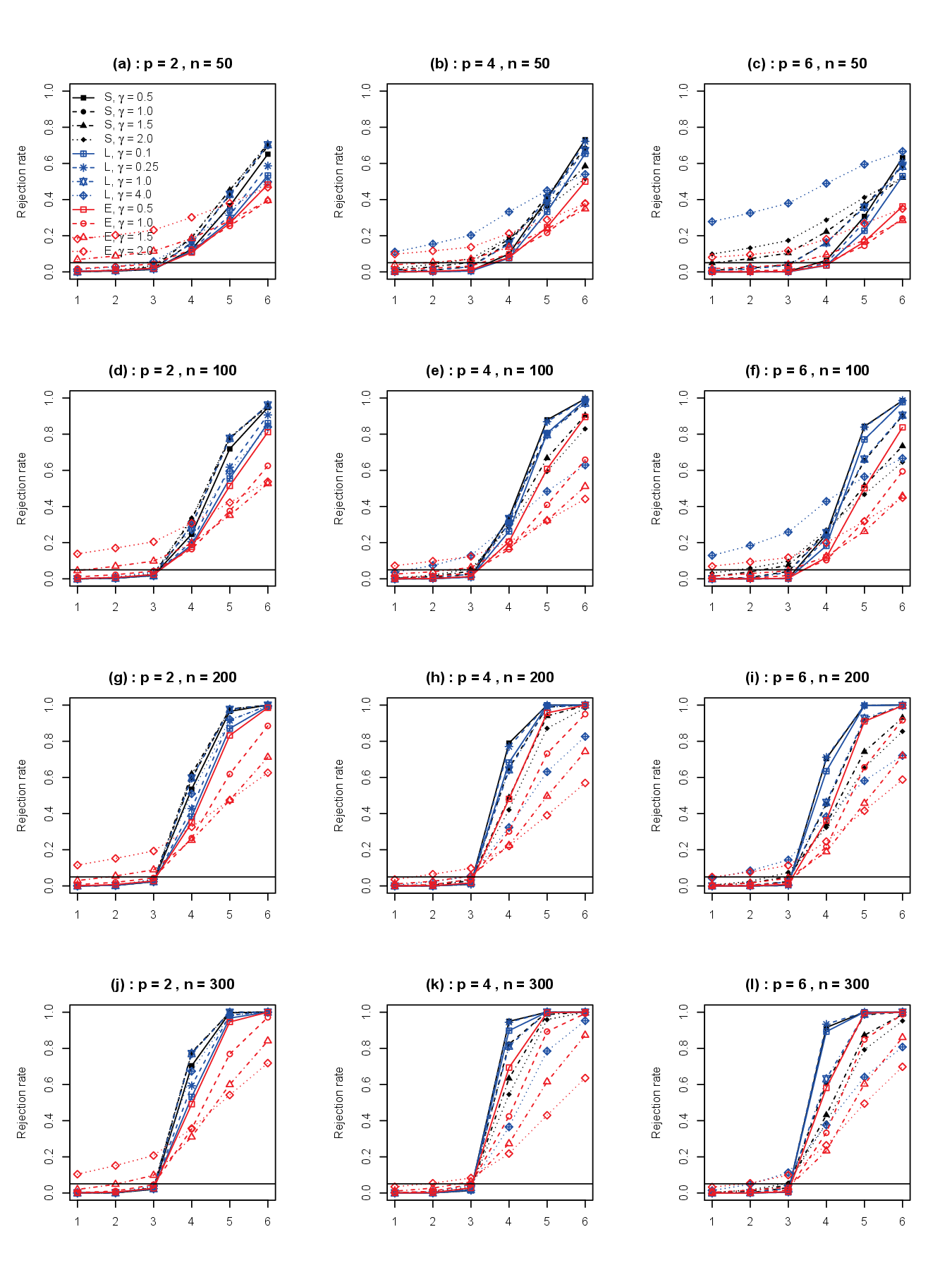}\\
\vspace{-1cm}
\caption{Simulation results for Example 3(b) based on $2000$ trials. The horizontal line corresponds to the $5\%$ significance level. The capital letters S, L, and E in panel (a) represent $T^{(\mathcal I)}_{n,S}$, $T^{(\mathcal I)}_{n,L}$, and $T^{(\mathcal I)}_{n,E}$ respectively.} \label{figI2}
\end{figure}

The reported empirical rejection rates for $T^{(\mathcal I)}_{n,S}$, $T^{(\mathcal I)}_{n,L}$ and $T^{(\mathcal I)}_{n,E}$ in Table \ref{tableI2} and Figures \ref{figI1}, \ref{figI2}, correspond to the benchmark value $\rho_0=0.8$ and the non-benchmark values employed $\rho \in \{0.84, 0.82, 0.8, 0.75, 0.7, 0.65\}$. Similar conclusions as before may be drawn from these results regarding Type--I errors and powers.
At the same time however, the energy--based test $T^{(\mathcal I)}_{n,E}$ seems to be less powerful for nearly all cases in Example 3(a) as well as in Example 3(b).

\begin{center}
  Table \ref{tableI2} and Figures \ref{figI1}, \ref{figI2} about here
\end{center}

The simulation results of Examples 1--3 suggest that in some cases the weight function used as well as the value of $\gamma$ may affect the performance of the proposed tests to a certain degree. Most research though on this topic concentrates on the choice of the weight parameter conditionally on the choice of the weight function, rather than on the choice of weight function itself. The hitherto most popular weight function has been the zero--mean normal density (or some variant thereof) and within the restricted context of goodness--of--fit testing for multivariate normality there is some work on how to choose $\gamma$. For this issue we refer to \cite{Henze1997, Tenreiro2009}, and the review of \cite{Meintanis2016}. In the more broad, though still strictly parametric,  context of testing goodness--of--fit to specific families of distributions \cite{Allison2015, Tenreiro2019} suggest data--driven ways for choosing the weight parameter. Now the more general issue of the proper choice of the particular functional form of the weight function has been investigated to some degree by \cite{Lindsay2014, Albert2022},  but the results are not readily applicable for actual test implementation in our setting. Instead here we opt for the practical approach of suggesting a compromise value of the weight parameter that seems to work well across several sampling scenarios. In this connection and based on our simulation results we suggest  the use of a ``medium" sized value of $\gamma$ such as $\gamma=1.0$ for both the spherical stable and the Laplace density as an overall good choice. Clearly though this suggestion is conditioned on our Monte Carlo results, and therefore can not be overgeneralized.

This section concludes with a numerical investigation of the random approximation procedure. In this connection, consider for instance the homogeneity--equivalence test in Example 2(a) and write the distance $\Delta^{(\mathcal H)}_{w,Z}$, with $X\sim \mathcal{N}_p(0,1)$ and $Y\sim \mathcal{N}_p(\mu_0,1)$, in the following form:
\begin{eqnarray} \label{InteH}
  \Delta^{(\mathcal H)}_{w,Z} &=& \int_{\mathbb R^p} \left|\varphi_X(t)-\varphi_Y(t)\right|^2 w(t){\rm{d}}t \\ \nonumber
   &=& \int_{\mathbb R^p} \left| \exp\left(-\frac{1}{2}t^\top t\right) - \exp\left(i\mu_0 1_p^\top t-\frac{1}{2}t^\top t\right) \right|^2 w(t){\rm{d}}t \\ \nonumber
   &=& 2 \int_{\mathbb R^p} \exp(-t^\top t) \left(1-\cos(\mu_01_p^\top t)\right)  w(t){\rm{d}}t,
\end{eqnarray}
(recall that $1_p$ denotes the $p$--dimensional vector of ones). 
Note in this connection that the density functions of the spherical stable distribution and the generalized spherical Laplace distribution can be found in \cite{Nolan2013} and \cite{Kozubowski2013}, respectively, and hence the values of $\Delta^{(\mathcal H)}_{w,Z}$ in Example 2(a) may be computed by numerical integration utilizing \eqref{InteH}.
The corresponding results reported in Table \ref{tableH1} are the values of $\Delta^{(\mathcal H)}_{w,Z}$ for the benchmark in Example 2(a) with $\mu_0=2.0$, first computed by random approximation and also by numerical integration, and show that the two methods are in close agreement, a fact that corroborates the use of the random approximation procedure.

\begin{center}
  Table~\ref{tableH1} about here
\end{center}

\section{A real--data application} \label{sec5}

In this section, we apply the proposed independence--equivalence test to a real--data set from AIDS Clinical Trials Group Protocol 175 (ACTG175), which is available from the \textsf{R} package \texttt{speff2trial} (\cite{Juraska2022}).
The proposed symmetry--equivalence and homogeneity-equivalence tests can be similarly applied in practice.
Note that the p--value corresponding to the critical region \eqref{CR} is
$$
  \Phi \left( \frac{\sqrt{n} \left( \Delta^{(\mathcal M)}_{n,w} - \Delta \right)}{\sigma_{n,w}^{(\mathcal M)}}  \right),
$$
where $\Phi(\cdot)$ denotes the cumulative distribution function of the standard normal distribution.
Due to good performance in the Monte Carlo study, we choose a ``medium" sized value of $\gamma=1.0$ for tests $T^{(\mathcal I)}_{n,S}$, $T^{(\mathcal I)}_{n,L}$, and $T^{(\mathcal I)}_{n,E}$ for simplicity.

ACTG175 was conducted as a randomized clinical trial in order to compare monotherapy with zidovudine or didanosine against combination therapy with zidovudine and didanosine or zidovudine and zalcitabine in adults infected with HIV.
The data set contains 2139 subjects, and the variable $arms$ indicates which treatment each subject was assigned to (0=zidovudine, 1=zidovudine and didanosine, 2=zidovudine and zalcitabine, 3=didanosine); see \cite{Hammer1996, Tsiatis2008} for more details.
\cite{Ma2023} recently performed independence tests under $arms=2$ group (524 subjects) between the treatment effect and some other covariates, and concluded that the null hypothesis of independence should be rejected at 5\% significance level due to the small p--values obtained.
Here the treatment effect was measured by the change of the CD4 count from baseline to 20 $\pm$ 5 weeks, and the covariates of interest are: history of intravenous drug use (0=no, 1=yes), antiretroviral history (0=naive, 1=experienced), age at baseline, and CD8 count at baseline.
Before testing the independence--equivalence of $X\in \mathbb R$ (the treatment effect) and $Y\in \mathbb R^4$ (the covariates of interest), we wish to point out that some dependence between treatment and covariates is actually expected due to the conclusion of \cite{Ma2023}, and thus the task of the independence--equivalence test is to assess the size of this dependence.
(All variables were standardized to have mean zero and variance one to eliminate the effect of measurement scale).

\begin{figure}[b]
\centering
\includegraphics[scale=0.7]{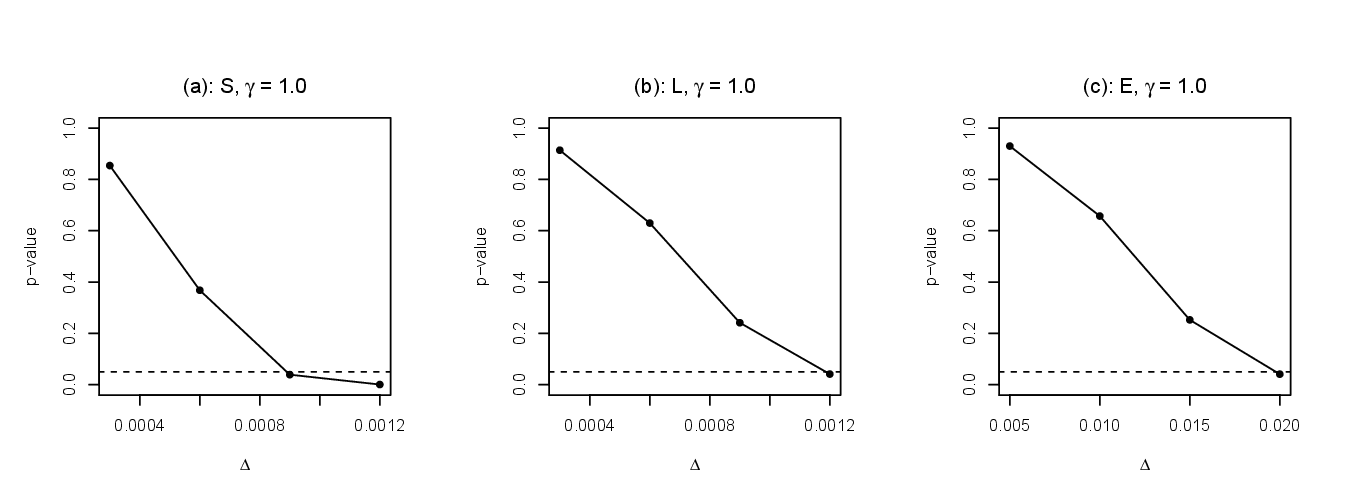}\\
\vspace{-0.5cm}
\caption{p--values of $T^{(\mathcal I)}_{n,S}$, $T^{(\mathcal I)}_{n,L}$, and $T^{(\mathcal I)}_{n,E}$ with $\gamma=1.0$ and different values of $\Delta$ for the analysis of ACTG175 data set. The horizontal dashed line corresponds to the $5\%$ significance level. } \label{figReal}
\end{figure}

To this end we chose four equally spaced values near the estimated $\Delta^{(\mathcal I)}_{n,w}$ as potential sizes of interest for the threshold $\Delta$.
 Table \ref{tableRealInde} shows the p-value corresponding to $T^{(\mathcal I)}_{n,S}$, $T^{(\mathcal I)}_{n,L}$, and $T^{(\mathcal I)}_{n,E}$ (with $\gamma=1.0$) for each such value of $\Delta$, while Figure \ref{figReal} depicts the p--value against different threshold size $\Delta$. As it may be inferred from the corresponding entry of Table \ref{tableRealInde} (with the spherical stable density as weight function), there is no evidence to reject  $\Delta^{(\mathcal I)}_{w,Z} \geq 0.0006$ at 5\% significance level (p--value 0.3681). On the other hand, the observed p--value= 0.0392 corresponding to the hypothesis $\Delta^{(\mathcal I)}_{w,Z} \geq 0.0009$ indicates a significant test result. Also looking at the left panel of Figure \ref{figReal} suggests that a $\Delta$--size of approximate magnitude just below 0.0009 may still be accepted, but the clearly visible decreasing p--value trend shows that this conclusion corresponds to a much lower level of confidence.
Analogous observations hold true for the criteria $T^{(\mathcal I)}_{n,L}$ and $T^{(\mathcal I)}_{n,E}$ corresponding to the Laplace weight function and the energy statistic, respectively. An overall conclusion appears to be that a certain amount of dependence is present between the treatment and covariates of the ACTG175 clinical trial, and that, in each case, the size of this dependence lies somewhat above the estimated value of $\Delta^{(\mathcal I)}_w$, but does not exceed   $\Delta^{(\mathcal I)}_{n,w}$ by a wide margin.

\begin{center}
  Table~\ref{tableRealInde} and Figure~\ref{figReal} about here
\end{center}

\section{Conclusions} \label{sec6}

In this paper, we propose three test criteria based on weighted $L_2$--type distance measures, each of which is appropriate for testing, respectively,  symmetry, distributional homogeneity, and  independence, all three hypotheses being formulated as equivalence hypotheses.
The asymptotic properties of the proposed tests are proved by utilizing the theory of $U$--statistics which  properly combined with certain choices of the adopted weighting scheme lead to computationally convenient, asymptotically normal tests that are consistent against any fixed alternative. Our Monte Carlo results and real--data analysis illustrate the applicability of the suggested procedures as well as their competitiveness vis--\'a--vis their energy--based counterparts. In perspective  we point out that the techniques employed herein may potentially be applied in the context of testing goodness--of--fit with arbitrary  parametric families of distributions.

%
%

\bibliographystyle{tfnlm}
\bibliography{neighbor}

\begin{thebibliography}{10}
\providecommand{\url}[1]{\normalfont{#1}}
\providecommand{\urlprefix}{Available from: }

\bibitem{Berger1987}
Berger~JO, Delampady~M. Testing precise hypotheses (with discussion). Statist
  Sci. 1987;\hspace{0pt}2(3):317--352.

\bibitem{Sprenger2013}
Sprenger~J. Testing a precise null hypothesis: the case of {L}indley's paradox.
  Philos Sci. 2013;\hspace{0pt}80(5):733--744.

\bibitem{Borovkov1998}
Borovkov~AA. Mathematical statistics. Gordon and Breach Science Publishers,
  Amsterdam; 1998. Translated from the Russian by A. Moullagaliev and revised
  by the author.

\bibitem{Wellek2010}
Wellek~S. Testing statistical hypotheses of equivalence and noninferiority. 2nd
  ed. CRC Press, Boca Raton, FL; 2010.

\bibitem{Dette2018}
Dette~H, M\"{o}llenhoff~K, Volgushev~S, et~al. Equivalence of regression
  curves. J Amer Statist Assoc. 2018;\hspace{0pt}113(522):711--729.

\bibitem{Dette1998}
Dette~H, Munk~A. Validation of linear regression models. Ann Statist.
  1998;\hspace{0pt}26(2):778--800.

\bibitem{Baringhaus2017}
Baringhaus~L, Ebner~B, Henze~N. The limit distribution of weighted
  {$L^2$}-goodness-of-fit statistics under fixed alternatives, with
  applications. Ann Inst Statist Math. 2017;\hspace{0pt}69(5):969--995.

\bibitem{Henze2020}
Henze~N, Mayer~C. More good news on the hkm test for multivariate reflected
  symmetry about an unknown centre. Ann Inst Statist Math.
  2020;\hspace{0pt}72:741--770.

\bibitem{Dette2021}
Dette~H, Kokot~K. Bio-equivalence tests in functional data by maximum
  deviation. Biometrika. 2021;\hspace{0pt}108(4):895--913.

\bibitem{Mollenhoff2022}
M\"{o}llenhoff~K, Dette~H, Bretz~F. Testing for similarity of binary
  efficacy-toxicity responses. Biostatistics. 2022;\hspace{0pt}23(3):949--966.

\bibitem{Chen2019}
Chen~F, Meintanis~SG, Zhu~L. On some characterizations and multidimensional
  criteria for testing homogeneity, symmetry and independence. J Multivariate
  Anal. 2019;\hspace{0pt}173:125--144.

\bibitem{Szekely2007}
Sz\'{e}kely~GJ, Rizzo~ML, Bakirov~NK. Measuring and testing dependence by
  correlation of distances. Ann Statist. 2007;\hspace{0pt}35(6):2769--2794.

\bibitem{Szekely2013}
Sz\'{e}kely~GJ, Rizzo~ML. Energy statistics: a class of statistics based on
  distances. J Statist Plann Inference. 2013;\hspace{0pt}143(8):1249--1272.

\bibitem{Nadarajah2003}
Nadarajah~S. The {K}otz-type distribution with applications. Statistics.
  2003;\hspace{0pt}37(4):341--358.

\bibitem{Kozubowski2013}
Kozubowski~TJ, Podg\'{o}rski~K, Rychlik~I. Multivariate generalized {L}aplace
  distribution and related random fields. J Multivariate Anal.
  2013;\hspace{0pt}113:59--72.

\bibitem{Hoeffding1948}
Hoeffding~W. A class of statistics with asymptotically normal distribution. Ann
  Math Statistics. 1948;\hspace{0pt}19:293--325.

\bibitem{Serfling1980}
Serfling~RJ. Approximation theorems of mathematical statistics. John Wiley \&
  Sons, Inc., New York; 1980. Wiley Series in Probability and Mathematical
  Statistics.

\bibitem{Lee1990}
Lee~AJ. U-statistics: Theory and practice. Taylor \& Francis Group, New York.;
  1990.

\bibitem{Efron1982}
Efron~B. The jackknife, the bootstrap and other resampling plans. (CBMS-NSF
  Regional Conference Series in Applied Mathematics; Vol.~38). Society for
  Industrial and Applied Mathematics (SIAM), Philadelphia, Pa.; 1982.

\bibitem{Quessy2012}
Quessy~JF, \'{E}thier~F. Cram\'{e}r-von {M}ises and characteristic function
  tests for the two and {$k$}-sample problems with dependent data. Comput
  Statist Data Anal. 2012;\hspace{0pt}56(6):2097--2111.

\bibitem{Hlavka2020}
Hl\'avka~Z, Hu\v{s}kov\'a~M, Meintanis~S. Change-point methods for multivariate
  time-series: paired vectorial observations. Statist Pap.
  2020;\hspace{0pt}61(4):1351--1383.

\bibitem{Nolan2013}
Nolan~JP. Multivariate elliptically contoured stable distributions: theory and
  estimation. Comput Statist. 2013;\hspace{0pt}28(5):2067--2089.

\bibitem{Gretton2012}
Gretton~A, Borgwardt~KM, Rasch~MJ, et~al. A kernel two-sample test. J Mach
  Learn Res. 2012;\hspace{0pt}13:723--773.

\bibitem{Micchelli2006}
Micchelli~C, Xu~Y, Zhang~H. Universal kernels. J Mach Learn Res.
  2006;\hspace{0pt}7:2651--2667.

\bibitem{Sejdinovic2013}
Sejdinovic~D, Sriperumbudur~B, Gretton~A, et~al. Equivalence of distance-based
  and {RKHS}-based statistics in hypothesis testing. Ann Statist.
  2013;\hspace{0pt}41(5):2263--2291.

\bibitem{Meintanis2012}
Meintanis~SG, Ngatchou-Wandji~J. Recent tests for symmetry with multivariate
  and structured data: a review. In: Nonparametric statistical methods and
  related topics. World Sci. Publ., Hackensack, NJ; 2012. p. 35--73.

\bibitem{Alba2008}
Alba~Fern\'{a}ndez~V, Jim\'{e}nez~Gamero~MD, Mu\~{n}oz Garc\'{i}a~J. A test for
  the two-sample problem based on empirical characteristic functions. Comput
  Statist Data Anal. 2008;\hspace{0pt}52(7):3730--3748.

\bibitem{Ghosh2016}
Ghosh~A, Biswas~M. Distribution--free high--dimensional two--sample tests based
  on discriminating hyperplanes. TEST. 2016;\hspace{0pt}25(3):525--547.

\bibitem{Alba2017}
Alba-Fern\'{a}ndez~MV, Batsidis~A, Jim\'{e}nez-Gamero~MD, et~al. A class of
  tests for the two-sample problem for count data. J Comput Appl Math.
  2017;\hspace{0pt}318:220--229.

\bibitem{Genest2004}
Genest~C, R\'emillard~B. Tests of independence and randomness based on the
  empirical copula process. TEST. 2004;\hspace{0pt}13(2):335--369.

\bibitem{Zhang2018}
Zhang~Q, Filippi~S, Gretton~A, et~al. Large--scale kernel methods for
  independence testing. Stat Comput. 2018;\hspace{0pt}28(1):113--130.

\bibitem{Azzalini2014}
Azzalini~A. The skew-normal and related families. (Institute of Mathematical
  Statistics (IMS) Monographs; Vol.~3). Cambridge University Press, Cambridge;
  2014. With the collaboration of Antonella Capitanio.

\bibitem{Azzalini2022}
Azzalini~A. The {R} package \texttt{sn}: The skew-normal and related
  distributions such as the skew-$t$ and the {SUN} (version 2.0.2).
  Universit\`a degli Studi di Padova, Italia; 2022. Home page:
  \url{http://azzalini.stat.unipd.it/SN/}.

\bibitem{Kotz2004}
Kotz~S, Nadarajah~S. Multivariate {$t$} distributions and their applications.
  Cambridge University Press, Cambridge; 2004.

\bibitem{Henze1997}
Henze~N. Extreme smoothing and testing for multivariate normality. Statist
  Probab Lett. 1997;\hspace{0pt}35:203--213.

\bibitem{Tenreiro2009}
Tenreiro~C. On the choice of the smoothing parameter for the {BHEP}
  goodness-of-fit test. Comput Statist Data Anal.
  2009;\hspace{0pt}53(4):1038--1053.

\bibitem{Meintanis2016}
Meintanis~SG. A review of testing procedures based on the empirical
  characteristic function (with discussion and rejoinder). South African
  Statist J. 2016;\hspace{0pt}50(1):1--14.

\bibitem{Allison2015}
Allison~JS, Santana~L. On a data-dependent choice of the tuning parameter
  appearing in certain goodness-of-fit tests. J Stat Comput Simul.
  2015;\hspace{0pt}85(16):3276--3288.

\bibitem{Tenreiro2019}
Tenreiro~C. On the automatic selection of the tuning parameter appearing in
  certain families of goodness-of-fit tests. J Stat Comput Simul.
  2019;\hspace{0pt}89(10):1780--1797.

\bibitem{Lindsay2014}
Lindsay~B, Markatou~M, Ray~S. Kernels, degrees of freedom, and power of
  quadratic distance goodness--of--fit tests. J Amer Statist Assoc.
  2014;\hspace{0pt}109:395--410.

\bibitem{Albert2022}
Albert~M, Laurent~B, Marrel~A, et~al. Adaptive test of independence based on
  hsic measures. Ann Statist. 2022;\hspace{0pt}50:858--879.

\bibitem{Juraska2022}
Michal~J, Gilbert~PB, Lu~X, et~al. The {R} package \texttt{speff2trial}:
  Semiparametric efficient estimation for a two-sample treatment effect
  (version 1.0.5); 2022. Home page:
  \url{https://CRAN.R-project.org/package=speff2trial}.

\bibitem{Hammer1996}
Hammer~SM, Katzenstein~DA, Hughes~MD, et~al. A trial comparing nucleoside
  monotherapy with combination therapy in hiv-infected adults with cd4 cell
  counts from 200 to 500 per cubic millimeter. New England Journal of Medicine.
  1996;\hspace{0pt}335(15):1081--1090. PMID: 8813038.

\bibitem{Tsiatis2008}
Tsiatis~AA, Davidian~M, Zhang~M, et~al. Covariate adjustment for two-sample
  treatment comparisons in randomized clinical trials: a principled yet
  flexible approach. Stat Med. 2008;\hspace{0pt}27(23):4658--4677.

\bibitem{Ma2023}
Ma~W, Ye~F, Xiao~J, et~al. A distribution-free test of independence based on a
  modified mean variance index. Statistical Theory and Related Fields.
  2023;\hspace{0pt}0(0):1--25.

\end{thebibliography}

\section*{Appendix}
We provide the technical proofs of Theorems \ref{thm}--ref{thm3} in this section.

\bigskip
\noindent\textbf{Proof of Theorem \ref{thm}.}
First, we proof the asymptotic properties of $\Delta^{(\mathcal S)}_{n,w}$ using the general theory of $U$--statistics.
Define $\psi_1^{\mathcal S}(x) = \mathbb E[\psi^{\mathcal S}(x,X)]$, where the expected value is taken with respect to the random vector $X$. Thus
\begin{eqnarray} \label{varS}
    Var[\psi_1^{\mathcal S}(X)]
    &=& \mathbb E\left[ \psi_1^{\mathcal S}(X) \right]^2 - \mathbb E^2 \left[\psi_1^{\mathcal S}(X)\right] \\ \nonumber
    &=& \mathbb E\left\{ \mathbb E[\psi^{\mathcal S}(X,X_1)|X] \right\}^2 - \mathbb E^2 \left\{ \mathbb E[\psi^{\mathcal S}(X,X_1)|X] \right\} \\ \nonumber
    &=& \mathbb E\left\{ \mathbb E[\psi^{\mathcal S}(X,X_1)\psi^{\mathcal S}(X,X_2)|X] \right\} - \mathbb E^2 \left[ \psi^{\mathcal S}(X,X_1) \right]  \\ \nonumber
    &=& \mathbb E\left[ \psi^{\mathcal S}(X,X_1)\psi^{\mathcal S}(X,X_2) \right] - \mathbb E^2 \left[ \psi^{\mathcal S}(X,X_1) \right],
\end{eqnarray}
where $X_1$ and $X_2$ are independent copies of $X$.
Consider that any characteristic function is uniformly bounded, thus ${\rm{Var}}[\psi_1^{\mathcal S}(X)] < \infty$ can be deduced without any extra assumptions about the distribution of $X$.
Hence according to the asymptotic property of $U$--statistics (\cite[see][\S 5.5.1]{Serfling1980}), as $n\rightarrow \infty$, we have
\begin{eqnarray*}
 \sqrt{n} \left(\Delta^{(\mathcal S)}_{n,w}-\Delta^{(\mathcal S)}_{w,X} \right)
 &=& \frac{2}{\sqrt{n}} \sum_{i=1}^n \left( \psi_1^{\mathcal S}(X_i) - \Delta^{(\mathcal S)}_{w,X} \right) + o_p(1) \\
 &\stackrel{{\cal D}}{\longrightarrow}& \mathcal N\left(0,\sigma_w^{2(\mathcal S)}\right),
\end{eqnarray*}
where $\sigma_w^{2(\mathcal S)} = 4 Var[\psi_1^{\mathcal S}(X)]$, and the limiting distribution is non--singular provided that $\sigma_w^{2(\mathcal S)} > 0$.
Notice further that $\sigma_w^{2(\mathcal S)} = 0$ if and only if the law of $X$ is symmetric about the origin, i.e., $\Delta^{(\mathcal S)}_{w,X} = 0$.
Thus when $\Delta^{(\mathcal S)}_{w,X} = 0$,
\begin{eqnarray*}
 \sqrt{n} \Delta^{(\mathcal S)}_{n,w} \stackrel{{\cal D}}{\longrightarrow} 0, \ \ as \ n\rightarrow \infty.
\end{eqnarray*}
The second part of the proof for $\Delta^{(\mathcal S)}_{n,w}$ can be finished immediately.

Second, we proof the asymptotic properties of $\Delta^{(\mathcal H)}_{n,w}$. Define $\psi_1^{\mathcal H}(z) = \mathbb E[\psi^{\mathcal H}(z,Z)]$, where the expected value is taken with respect to the random vector $Z$.
Similar to \eqref{varS}, we have
\begin{eqnarray*} \label{varH}
    Var[\psi_1^{\mathcal H}(Z)]
    &=& \mathbb E\left[ \psi^{\mathcal H}(Z,Z_1)\psi^{\mathcal H}(Z,Z_2) \right] - \mathbb E^2 \left[ \psi^{\mathcal H}(Z,Z_1) \right].
\end{eqnarray*}
Then the asymptotic properties of $\Delta^{(\mathcal H)}_{n,w}$ can be obtained by similar arguments as in the case of $\Delta^{(\mathcal S)}_{n,w}$.

Now we proof the asymptotic properties of $\Delta^{(\mathcal I)}_{n,w}$. Define $\psi_1^{\mathcal I,i}(z) = \mathbb E[\psi^{\mathcal I,i}(z,Z_1,\ldots,Z_{m_i-1})]$, $i=1,\ldots,4$, where $m_i$ is the degree of kernel $\psi^{\mathcal I,i}$, and the expected value is taken with respect to the random vectors $Z_1,\ldots,Z_{m_i-1}$.
Further, define
\begin{eqnarray*}
  \zeta_1(i,j) &=& Cov\left( \psi_1^{\mathcal I,i}(Z),\psi_1^{\mathcal I,j}(Z) \right), \ \ i,j=1,\ldots,4.
\end{eqnarray*}
Obviously $\zeta_1(i,j)$ is symmetric in the sense that $\zeta_1(i,j)=\zeta_1(j,i)$.
Similar to \eqref{varS}, for $i,j = 1,2,3$, we have
\begin{eqnarray*}
& & \zeta_1(i,j) = \mathbb E\left[ \psi^{\mathcal I,i}(Z,Z_1)\psi^{\mathcal I,j}(Z,Z_2) \right] - S_iS_j, \\
& & \zeta_1(i,4) = \mathbb E\left[ \psi^{\mathcal I,i}(Z,Z_1)\psi^{\mathcal I,4}(Z,Z_2,Z_3) \right] - S_iS_4, \\
& & \zeta_1(4,4) = \mathbb E\left[ \psi^{\mathcal I,4}(Z,Z_1,Z_2)\psi^{\mathcal I,4}(Z,Z_3,Z_4) \right] - S_4^2.
\end{eqnarray*}
Here all pairs of kernels $\psi^{\mathcal I,i}$ and $\psi^{\mathcal I,j}$ are evaluated on two subsets with only one element in common.
Thus $\zeta_1(i,j) < \infty$, $i,j=1,\ldots,4$, can be obtained according to the uniform boundedness of characteristic functions, and the proof can be finished by using Theorem 7.5 in \cite{Hoeffding1948} and the fact that $\sigma_{w}^{2(\mathcal I)}=0$ holds true if and only if $X$ and $Y$ are mutually independent.

\bigskip
\noindent\textbf{Proof of Theorem \ref{thm2}.}
Note that the SLLN for $U$--statistics (\cite[see][Theorem 5.4A]{Serfling1980}) applies under the condition that the corresponding kernel has a bounded expectation.
In this connection, kernels associated with $\sigma_{n,w}^{2(\mathcal S)}$ and $\sigma_{n,w}^{2(\mathcal H)}$ only depend on $C_w(\cdot)$, and thus in our context this condition translates into the condition that $\mathbb E|C_w(\cdot)|^2<\infty$, which is trivially true for any real--valued CF and any underlying law of observations.
Then the conclusion readily follows by invoking the SLLN for $U$--statistics and Slutsky's theorem.

\bigskip
\noindent\textbf{Proof of Theorem \ref{thm3}.}
Recall that $\Delta^{(\mathcal I)}_{n,w}$ is a function of the $U$--statistics $U_{n,1},\ldots,U_{n,4}$.
Thus the proof of the theorem follows easily by using the theory of jackknife functions of several $U$--statistics; see \S 5.1.3 in \cite{Lee1990}.

\newpage
\begin{table}[p]
  \centering
  \caption{Simulation results for Example 1. Rejection rates of $T^{(\mathcal S)}_{n,S}$ at $5\%$ significance level using $2000$ trials. } \label{tableS2}
  \renewcommand\arraystretch{1.2}
  \scalebox{0.58}{
    \begin{tabular}{cccccccccccccccccccccccccccccccccc}
    \cline{1-18}
   &   &   &\multicolumn{7}{c}{Example 1(a)}   &   &\multicolumn{7}{c}{Example 1(b)}
   \\    \cline{4-10}    \cline{12-18}
   &   &   &\multicolumn{3}{c}{Null hypotheses}    &    &\multicolumn{3}{c}{Alternatives}   &   &\multicolumn{3}{c}{Null hypotheses}    &    &\multicolumn{3}{c}{Alternatives}
   \\    \cline{4-6}    \cline{8-10}    \cline{12-14}    \cline{16-18}
   &   &   &$\theta=5$  &$\theta=4$  &$\theta=3$  &  &$\theta=2$  &$\theta=1$  &$\theta=0$  &   &$\theta=5$  &$\theta=4$  &$\theta=3$  &  &$\theta=2$  &$\theta=1$  &$\theta=0$
   \\    \cline{1-18}
$n=50$ &$p=2$ &$\gamma=0.5$   &0.0440 	&0.0520 	&0.0765  &	&0.1970 	&0.6740 	&1      	&	&0.0615 	&0.0790 	&0.1195 	&	&0.2245 	&0.6690 	&1\\
      &      &$\gamma=1.0$    &0.0425 	&0.0550 	&0.0855  &	&0.2080 	&0.6910 	&1 	        &	&0.0780 	&0.0935 	&0.1325 	&	&0.2370 	&0.6495 	&1\\
      &      &$\gamma=1.5$    &0.0465 	&0.0630 	&0.0995  &	&0.2225 	&0.7135 	&1 	        &	&0.0875 	&0.1075 	&0.1410 	&	&0.2500 	&0.6475 	&1\\
      &      &$\gamma=2.0$    &0.0480 	&0.0645 	&0.1065  &	&0.2275 	&0.7315 	&1 	        &	&0.0975 	&0.1165 	&0.1515 	&	&0.2590 	&0.6500 	&1\\
    \cline{2-18}															
      &$p=4$ &$\gamma=0.5$    &0.0865 	&0.0905 	&0.0995  &	&0.1395 	&0.3545 	&1 	        &	&0.0875 	&0.0920 	&0.1040 	&	&0.1380 	&0.3180 	&0.9995\\
      &      &$\gamma=1.0$    &0.0860 	&0.0905 	&0.1050  &	&0.1395 	&0.3780 	&1 	        &	&0.1125 	&0.1155 	&0.1255 	&	&0.1550 	&0.3255 	&0.9935\\
      &      &$\gamma=1.5$    &0.0970 	&0.1030 	&0.1220  &	&0.1610 	&0.4105 	&1      	&	&0.1400 	&0.1455 	&0.1615 	&	&0.1965 	&0.3570 	&0.9770\\
      &      &$\gamma=2.0$    &0.1160 	&0.1220 	&0.1365  &	&0.1790 	&0.4430 	&0.9990 	&	&0.1710 	&0.1800 	&0.1890 	&	&0.2210 	&0.3965 	&0.9670\\
    \cline{2-18}															
      &$p=6$ &$\gamma=0.5$    &0.1145 	&0.1180 	&0.1265  &	&0.1550 	&0.2960 	&1 	        &	&0.1345 	&0.1385 	&0.1435 	&	&0.1600 	&0.2640 	&0.9980\\
      &      &$\gamma=1.0$    &0.1200 	&0.1220 	&0.1340  &	&0.1590 	&0.3020 	&0.9995 	&	&0.1710 	&0.1720 	&0.1785 	&	&0.1975 	&0.2975 	&0.9710\\
      &      &$\gamma=1.5$    &0.1470 	&0.1525 	&0.1615  &	&0.1880 	&0.3225 	&0.9960 	&	&0.2220 	&0.2265 	&0.2315 	&	&0.2440 	&0.3460 	&0.9210\\
      &      &$\gamma=2.0$    &0.1790 	&0.1840 	&0.1965  &	&0.2360 	&0.3710 	&0.9850 	&	&0.2490 	&0.2525 	&0.2560 	&	&0.2780 	&0.3685 	&0.8740\\
    \cline{1-18}
$n=100$ &$p=2$ &$\gamma=0.5$  &0.0185 	&0.0300 	&0.0575  & 	&0.1890 	&0.8615 	&1  &	&0.0385 	&0.0520 	&0.0920 	&	&0.2245 	&0.8320 	&1\\
      &      &$\gamma=1.0$    &0.0200 	&0.0315 	&0.0585  & 	&0.2020 	&0.8755 	&1 	&	&0.0465 	&0.0610 	&0.0945 	&	&0.2315 	&0.8095 	&1\\
      &      &$\gamma=1.5$    &0.0215 	&0.0320 	&0.0620  & 	&0.2100 	&0.8870 	&1 	&	&0.0495 	&0.0620 	&0.1080 	&	&0.2410 	&0.7975 	&1\\
      &      &$\gamma=2.0$    &0.0235 	&0.0330 	&0.0645  & 	&0.2235 	&0.8930 	&1 	&	&0.0495 	&0.0675 	&0.1130 	&	&0.2465 	&0.7960 	&1\\
    \cline{2-18}
      &$p=4$ &$\gamma=0.5$    &0.0410 	&0.0460 	&0.0580  & 	&0.1070 	&0.4590 	&1 	&	&0.0405 	&0.0440 	&0.0545 	&	&0.0950 	&0.3855 	&1\\
      &      &$\gamma=1.0$    &0.0430 	&0.0450 	&0.0570  & 	&0.1130 	&0.4835 	&1 	&	&0.0565 	&0.0625 	&0.0770 	&	&0.1090 	&0.3480 	&1\\
      &      &$\gamma=1.5$    &0.0430 	&0.0500 	&0.0630  & 	&0.1200 	&0.5120 	&1 	&	&0.0755 	&0.0810 	&0.0910 	&	&0.1335 	&0.3550 	&1\\
      &      &$\gamma=2.0$    &0.0505 	&0.0550 	&0.0700  & 	&0.1310 	&0.5400 	&1 	&	&0.0870 	&0.0925 	&0.1060 	&	&0.1480 	&0.3735 	&0.9985\\
\cline{2-18}
      &$p=6$ &$\gamma=0.5$    &0.0785 	&0.0840 	&0.0985  & 	&0.1285 	&0.3410 	&1 	&	&0.0815 	&0.0855 	&0.0910 	&	&0.1135 	&0.2705 	&1\\
      &      &$\gamma=1.0$    &0.0790 	&0.0830 	&0.0975  & 	&0.1335 	&0.3460 	&1 	&	&0.0860 	&0.0880 	&0.0940 	&	&0.1140 	&0.2470 	&0.9990\\
      &      &$\gamma=1.5$    &0.0815 	&0.0855 	&0.0950  & 	&0.1320 	&0.3535 	&1 	&	&0.1060 	&0.1085 	&0.1225 	&	&0.1415 	&0.2725 	&0.9920\\
      &      &$\gamma=2.0$    &0.0990 	&0.1065 	&0.1205  & 	&0.1465 	&0.3770 	&1 	&	&0.1420 	&0.1435 	&0.1515 	&	&0.1745 	&0.2955 	&0.9765\\
    \cline{1-18}										
$n=200$ &$p=2$ &$\gamma=0.5$  &0.0085 	&0.0145 	&0.0430  & 	&0.2575 	&0.9825 	&1 	&	&0.0165 	&0.0255 	&0.0705 	&	&0.2765 	&0.9705 	&1\\
      &      &$\gamma=1.0$    &0.0100 	&0.0160 	&0.0500  & 	&0.2730 	&0.9865 	&1 	&	&0.0175 	&0.0330 	&0.0800 	&	&0.2930 	&0.9565 	&1\\
      &      &$\gamma=1.5$    &0.0105 	&0.0180 	&0.0590  & 	&0.2940 	&0.9900 	&1 	&	&0.0200 	&0.0385 	&0.0900 	&	&0.3005 	&0.9450 	&1\\
      &      &$\gamma=2.0$    &0.0105 	&0.0210 	&0.0670  & 	&0.3150 	&0.9900 	&1 	&	&0.0250 	&0.0440 	&0.0960 	&	&0.3025 	&0.9375 	&1\\
    \cline{2-18}
      &$p=4$ &$\gamma=0.5$    &0.0230 	&0.0280 	&0.0415  & 	&0.1035 	&0.6525 	&1 	&	&0.0165 	&0.0225 	&0.0300 	&	&0.0680 	&0.4965 	&1\\
      &      &$\gamma=1.0$    &0.0220 	&0.0275 	&0.0430  & 	&0.1045 	&0.6855 	&1 	&	&0.0230 	&0.0275 	&0.0355 	&	&0.0790 	&0.4405 	&1\\
      &      &$\gamma=1.5$    &0.0235 	&0.0315 	&0.0480  & 	&0.1185 	&0.7145 	&1 	&	&0.0345 	&0.0390 	&0.0530 	&	&0.0985 	&0.4420 	&1\\
      &      &$\gamma=2.0$    &0.0300 	&0.0365 	&0.0540  & 	&0.1350 	&0.7285 	&1 	&	&0.0425 	&0.0505 	&0.0685 	&	&0.1130 	&0.4530 	&1\\
    \cline{2-18}
      &$p=6$ &$\gamma=0.5$    &0.0540 	&0.0615 	&0.0750  & 	&0.1090 	&0.4485 	&1 	&	&0.0465 	&0.0495 	&0.0540 	&	&0.0870 	&0.3245 	&1\\
      &      &$\gamma=1.0$    &0.0490 	&0.0550 	&0.0680  & 	&0.1055 	&0.4555 	&1 	&	&0.0455 	&0.0485 	&0.0535 	&	&0.0810 	&0.2710 	&1\\
      &      &$\gamma=1.5$    &0.0460 	&0.0530 	&0.0645  & 	&0.1015 	&0.4285 	&1 	&	&0.0580 	&0.0615 	&0.0675 	&	&0.0955 	&0.2595 	&1\\
      &      &$\gamma=2.0$    &0.0510 	&0.0570 	&0.0630  & 	&0.1020 	&0.4315 	&1 	&	&0.0720 	&0.0750 	&0.0845 	&	&0.1105 	&0.2690 	&1\\
    \cline{1-18}
$n=300$ &$p=2$ &$\gamma=0.5$  &0.0040 	&0.0120 	&0.0425  & 	&0.3270 	&1 	 	    &1 	&	&0.0070 	&0.0225 	&0.0705 	&	&0.3800 	&0.9960 	&1\\
      &      &$\gamma=1.0$    &0.0045 	&0.0145 	&0.0520  & 	&0.3455 	&1 	 	    &1 	&	&0.0155 	&0.0265 	&0.0795 	&	&0.3660 	&0.9940 	&1\\
      &      &$\gamma=1.5$    &0.0055 	&0.0140 	&0.0600  & 	&0.3715 	&1 	 	    &1 	&	&0.0165 	&0.0310 	&0.0825 	&	&0.3765 	&0.9910 	&1\\
      &      &$\gamma=2.0$    &0.0055 	&0.0135 	&0.0645  & 	&0.3915 	&1 	 	    &1 	&	&0.0180 	&0.0320 	&0.0900 	&	&0.3760 	&0.9895 	&1\\
    \cline{2-18}
      &$p=4$ &$\gamma=0.5$    &0.0205 	&0.0265 	&0.0405  & 	&0.1135 	&0.8065 	&1 	&	&0.0115 	&0.0155 	&0.0280 	&	&0.0695 	&0.6190 	&1\\
      &      &$\gamma=1.0$    &0.0200 	&0.0255 	&0.0390  & 	&0.1160 	&0.8290 	&1 	&	&0.0155 	&0.0210 	&0.0290 	&	&0.0760 	&0.5170 	&1\\
      &      &$\gamma=1.5$    &0.0175 	&0.0245 	&0.0430  & 	&0.1260 	&0.8510 	&1 	&	&0.0235 	&0.0300 	&0.0420 	&	&0.0885 	&0.5180 	&1\\
      &      &$\gamma=2.0$    &0.0185 	&0.0255 	&0.0460  & 	&0.1400 	&0.8645 	&1 	&	&0.0310 	&0.0365 	&0.0515 	&	&0.1065 	&0.5335 	&1\\
    \cline{2-18}
      &$p=6$ &$\gamma=0.5$    &0.0430 	&0.0490 	&0.0655  & 	&0.1210 	&0.5815 	&1 	&	&0.0285 	&0.0295 	&0.0355 	&	&0.0605 	&0.3735 	&1\\
      &      &$\gamma=1.0$    &0.0395 	&0.0415 	&0.0515  & 	&0.1080 	&0.5820 	&1 	&	&0.0275 	&0.0305 	&0.0355 	&	&0.0575 	&0.2815 	&1\\
      &      &$\gamma=1.5$    &0.0310 	&0.0340 	&0.0445  & 	&0.0900 	&0.5485 	&1 	&	&0.0385 	&0.0415 	&0.0490 	&	&0.0705 	&0.2625 	&1\\
      &      &$\gamma=2.0$    &0.0295 	&0.0345 	&0.0465  & 	&0.0895 	&0.5400 	&1 	&	&0.0505 	&0.0540 	&0.0600 	&	&0.0815 	&0.2625 	&1\\
    \cline{1-18}
  \end{tabular}
  }
\end{table}%

\newpage
\begin{table}[p]
  \centering
  \caption{Simulation results for Example 2.  Rejection rates of $T^{(\mathcal H)}_{n,S}$ at $5\%$ significance level using $2000$ trials. } \label{tableH2}
  \renewcommand\arraystretch{1.2}
  \scalebox{0.58}{
    \begin{tabular}{cccccccccccccccccccccccccccccccccc}
    \cline{1-18}
   &   &   &\multicolumn{7}{c}{Example 2(a)}   &   &\multicolumn{7}{c}{Example 2(b)}
   \\    \cline{4-10}    \cline{12-18}
   &   &   &\multicolumn{3}{c}{Null hypotheses}    &    &\multicolumn{3}{c}{Alternatives}   &   &\multicolumn{3}{c}{Null hypotheses}    &    &\multicolumn{3}{c}{Alternatives}
   \\    \cline{4-6}    \cline{8-10}    \cline{12-14}    \cline{16-18}
   &   &   &$\mu=2.2$  &$\mu=2.1$  &$\mu=2.0$  &  &$\mu=1.9$  &$\mu=1.8$  &$\mu=1.7$  &   &$\mu=2.2$  &$\mu=2.1$  &$\mu=2.0$  &  &$\mu=1.9$  &$\mu=1.8$  &$\mu=1.7$
   \\    \cline{1-18}
$n=50$ &$p=2$ &$\gamma=0.5$   &0.0025 	&0.0160 	&0.0525 	&	&0.1375 	&0.2935 	&0.4960 	&	&0.0015 	&0.0220 	&0.0735 	&	&0.2015 	&0.4565 	&0.7425\\
      &      &$\gamma=1.0$    &0.0075 	&0.0205 	&0.0535 	&	&0.1200 	&0.2370 	&0.3825 	&	&0.0025 	&0.0190 	&0.0760 	&	&0.2175 	&0.4655 	&0.7430\\
      &      &$\gamma=1.5$    &0.0135 	&0.0280 	&0.0560 	&	&0.1085 	&0.1950 	&0.2940 	&	&0.0045 	&0.0250 	&0.0750 	&	&0.2275 	&0.4680 	&0.7315\\
      &      &$\gamma=2.0$    &0.0195 	&0.0335 	&0.0610 	&	&0.1050 	&0.1810 	&0.2690 	&	&0.0055 	&0.0245 	&0.0765 	&	&0.2375 	&0.4775 	&0.7310\\
    \cline{2-18}														
      &$p=4$ &$\gamma=0.5$    &0.0015 	&0.0120 	&0.0595 	&	&0.1930 	&0.4235 	&0.7005 	&	&0.0005 	&0.0070 	&0.0545 	&	&0.2805 	&0.6780 	&0.9395\\
      &      &$\gamma=1.0$    &0.0170 	&0.0375 	&0.0695 	&	&0.1315 	&0.2480 	&0.3785 	&	&0.0010 	&0.0090 	&0.0635 	&	&0.2990 	&0.6795 	&0.9275\\
      &      &$\gamma=1.5$    &0.0595 	&0.0740 	&0.0925 	&	&0.1215 	&0.1635 	&0.2230 	&	&0.0020 	&0.0195 	&0.0960 	&	&0.3340 	&0.6755 	&0.9040\\
      &      &$\gamma=2.0$    &0.0915 	&0.1005 	&0.1140 	&	&0.1330 	&0.1615 	&0.1935 	&	&0.0045 	&0.0285 	&0.1245 	&	&0.3655 	&0.6710 	&0.8900\\
    \cline{2-18}														
      &$p=6$ &$\gamma=0.5$    &0.0010 	&0.0120 	&0.0640 	&	&0.2220 	&0.5195 	&0.8215 	&	&0.0000 	&0.0015 	&0.0505 	&	&0.3800 	&0.8225 	&0.9895\\
      &      &$\gamma=1.0$    &0.0335 	&0.0515 	&0.0850 	&	&0.1325 	&0.2140 	&0.3255 	&	&0.0000 	&0.0095 	&0.0865 	&	&0.3855 	&0.7915 	&0.9750\\
      &      &$\gamma=1.5$    &0.1005 	&0.1060 	&0.1105 	&	&0.1190 	&0.1390 	&0.1650 	&	&0.0030 	&0.0240 	&0.1345 	&	&0.4365 	&0.7750 	&0.9535\\
      &      &$\gamma=2.0$    &0.1470 	&0.1475 	&0.1520 	&	&0.1565 	&0.1655 	&0.1820 	&	&0.0125 	&0.0550 	&0.2070 	&	&0.4850 	&0.7745 	&0.9350\\
    \cline{1-18}
$n=100$ &$p=2$ &$\gamma=0.5$  &0.0010 	&0.0080 	&0.0510 	& 	&0.1810 	&0.4285 	&0.7090 &	&0.0005 	&0.0075 	&0.0545 	&	&0.2580 	&0.6475 	&0.9215\\
      &      &$\gamma=1.0$    &0.0020 	&0.0120 	&0.0500 	& 	&0.1390 	&0.3230 	&0.5550 &	&0.0005 	&0.0075 	&0.0570 	&	&0.2655 	&0.6500 	&0.9175\\
      &      &$\gamma=1.5$    &0.0050 	&0.0185 	&0.0510 	& 	&0.1195 	&0.2490 	&0.4240 &	&0 	&0.0080 	&0.0565 	&	&0.2675 	&0.6475 	&0.9140\\
      &      &$\gamma=2.0$    &0.0070 	&0.0215 	&0.0525 	& 	&0.1090 	&0.2120 	&0.3735 &	&0 	&0.0075 	&0.0605 	&	&0.2750 	&0.6410 	&0.9130\\
    \cline{2-18}
      &$p=4$ &$\gamma=0.5$    &0     	&0.0030 	&0.0500 	& 	&0.2490 	&0.6325 	&0.9070 &	&0 	&0.0005 	&0.0525 	&	&0.3955 	&0.8840 	&0.9980\\
      &      &$\gamma=1.0$    &0.0075 	&0.0210 	&0.0565 	& 	&0.1515 	&0.3060 	&0.5395 &	&0 	&0.0010 	&0.0660 	&	&0.4020 	&0.8785 	&0.9950\\
      &      &$\gamma=1.5$    &0.0335 	&0.0450 	&0.0655 	& 	&0.0995 	&0.1575 	&0.2560 &	&0 	&0.0030 	&0.0825 	&	&0.4120 	&0.8480 	&0.9880\\
      &      &$\gamma=2.0$    &0.0510 	&0.0585 	&0.0740 	& 	&0.0925 	&0.1340 	&0.1860 &	&0 	&0.0085 	&0.0960 	&	&0.4220 	&0.8360 	&0.9835\\
    \cline{2-18}
      &$p=6$ &$\gamma=0.5$    &0    	&0.0030 	&0.0495 	& 	&0.3225 	&0.7675 	&0.9790 &	&0 	&0 	&0.0400 	&	&0.5260 	&0.9655 	&1\\
      &      &$\gamma=1.0$    &0.0115 	&0.0270 	&0.0570 	& 	&0.1275 	&0.2580 	&0.4450 &	&0 	&0 	&0.0545 	&	&0.5085 	&0.9405 	&1\\
      &      &$\gamma=1.5$    &0.0625 	&0.0695 	&0.0755 	& 	&0.0895 	&0.1150 	&0.1485 &	&0 	&0.0030 	&0.0795 	&	&0.5100 	&0.9180 	&0.9975\\
      &      &$\gamma=2.0$    &0.0955 	&0.0980 	&0.1035 	& 	&0.1070 	&0.1155 	&0.1390 &	&0 	&0.0065 	&0.1135 	&	&0.5345 	&0.9035 	&0.9940\\
    \cline{1-18}
$n=200$ &$p=2$ &$\gamma=0.5$  &0    	&0.0010 	&0.0450 	& 	&0.2560 	&0.6540 	&0.9315 &	&0 	&0 	&0.0515 	&	&0.3795 	&0.8600 	&0.9985\\
      &      &$\gamma=1.0$    &0    	&0.0035 	&0.0400 	& 	&0.1940 	&0.4870 	&0.8055 &	&0 	&0 	&0.0575 	&	&0.3935 	&0.8690 	&0.9960\\
      &      &$\gamma=1.5$    &0    	&0.0060 	&0.0420 	& 	&0.1480 	&0.3570 	&0.6250 &	&0 	&0 	&0.0570 	&	&0.3920 	&0.8615 	&0.9950\\
      &      &$\gamma=2.0$    &0.0015 	&0.0085 	&0.0405 	& 	&0.1275 	&0.3010 	&0.5405 &	&0 	&0.0005 	&0.0570 	&	&0.3885 	&0.8575 	&0.9935\\
    \cline{2-18}
      &$p=4$ &$\gamma=0.5$    &0    	&0.0015 	&0.0430 	& 	&0.3845 	&0.8740 	&0.9925 &	&0 	&0 	&0.0450 	&	&0.6055 	&0.9920 	&1\\
      &      &$\gamma=1.0$    &0.0015 	&0.0080 	&0.0495 	& 	&0.1750 	&0.4625 	&0.7780 &	&0 	&0 	&0.0490 	&	&0.6010 	&0.9880 	&1\\
      &      &$\gamma=1.5$    &0.0170 	&0.0280 	&0.0530 	& 	&0.0980 	&0.1910 	&0.3425 &	&0 	&0 	&0.0565 	&	&0.5800 	&0.9800 	&1\\
      &      &$\gamma=2.0$    &0.0295 	&0.0405 	&0.0595 	& 	&0.0860 	&0.1350 	&0.2265 &	&0 	&0.0005 	&0.0590 	&	&0.5690 	&0.9700 	&1\\
    \cline{2-18}
      &$p=6$ &$\gamma=0.5$    &0    	&0      	&0.0495 	& 	&0.5100 	&0.9570 	&0.9995 &	&0 	&0 	&0.0400 	&	&0.7715 	&0.9990 	&1\\
      &      &$\gamma=1.0$    &0.0075 	&0.0185 	&0.0520 	& 	&0.1580 	&0.3700 	&0.6505 &	&0 	&0 	&0.0510 	&	&0.7145 	&0.9980 	&1\\
      &      &$\gamma=1.5$    &0.0395 	&0.0485 	&0.0600 	& 	&0.0805 	&0.1100 	&0.1625 &	&0 	&0 	&0.0645 	&	&0.6915 	&0.9945 	&1\\
      &      &$\gamma=2.0$    &0.0635 	&0.0670 	&0.0725 	& 	&0.0785 	&0.0930 	&0.1115 &	&0 	&0.0010 	&0.0860 	&	&0.6925 	&0.9885 	&1\\
    \cline{1-18}
$n=300$ &$p=2$ &$\gamma=0.5$  &0    	&0.0010 	&0.0390 	& 	&0.3245 	&0.8050 	&0.9890 &	&0 	&0 	&0.0405 	&	&0.4815 	&0.9520 	&1\\
      &      &$\gamma=1.0$    &0    	&0.0025 	&0.0340 	& 	&0.2280 	&0.6375 	&0.9260 &	&0 	&0.0005 	&0.0475 	&	&0.4905 	&0.9555 	&1\\
      &      &$\gamma=1.5$    &0    	&0.0040 	&0.0325 	& 	&0.1625 	&0.4565 	&0.7910 &	&0 	&0.0005 	&0.0475 	&	&0.4835 	&0.9465 	&1\\
      &      &$\gamma=2.0$    &0.0005 	&0.0060 	&0.0305 	& 	&0.1335 	&0.3850 	&0.6990 &	&0 	&0.0005 	&0.0475 	&	&0.4775 	&0.9410 	&1\\
    \cline{2-18}						
      &$p=4$ &$\gamma=0.5$    &0    	&0      	&0.0445 	& 	&0.5050 	&0.9600 	&0.9995 &	&0 	&0 	&0.0405 	&	&0.7565 	&1 	&1\\
      &      &$\gamma=1.0$    &0.0005 	&0.0080 	&0.0415 	& 	&0.2200 	&0.5815 	&0.8945 &	&0 	&0 	&0.0480 	&	&0.7510 	&1 	&1\\
      &      &$\gamma=1.5$    &0.0115 	&0.0235 	&0.0480 	& 	&0.1090 	&0.2260 	&0.4240 &	&0 	&0 	&0.0475 	&	&0.7200 	&0.9980 	&1\\
      &      &$\gamma=2.0$    &0.0215 	&0.0305 	&0.0515 	& 	&0.0895 	&0.1550 	&0.2700 &	&0 	&0 	&0.0535 	&	&0.6995 	&0.9960 	&1\\						
    \cline{2-18}
      &$p=6$ &$\gamma=0.5$    &0    	&0.0005 	&0.0415 	& 	&0.6205 	&0.9920 	&1 	&	&0 	&0 	&0.0280 	&	&0.8925 	&1 	&1\\
      &      &$\gamma=1.0$    &0.0020 	&0.0115 	&0.0450 	& 	&0.1690 	&0.4485 	&0.7850 &	&0 	&0 	&0.0460 	&	&0.8345 	&1 	&1\\
      &      &$\gamma=1.5$    &0.0295 	&0.0350 	&0.0450 	& 	&0.0755 	&0.1170 	&0.1735 &	&0 	&0 	&0.0615 	&	&0.7910 	&0.9995 	&1\\
      &      &$\gamma=2.0$    &0.0470 	&0.0515 	&0.0585 	& 	&0.0670 	&0.0855 	&0.1090 &	&0 	&0 	&0.0745 	&	&0.7755 	&0.9990 	&1\\
    \cline{1-18}
  \end{tabular}
  }
\end{table}%

\newpage
\begin{table}[p]
  \centering
  \caption{Simulation results for Example 3. Rejection rates of $T^{(\mathcal I)}_{n,S}$ at $5\%$ significance level using $2000$ trials with the variance of $\Delta^{(\mathcal I)}_{n,w}$ estimated by jackknife.} \label{tableI2}
  \renewcommand\arraystretch{1.18}
  \scalebox{0.58}{
    \begin{tabular}{cccccccccccccccccccccccccccccccccc}
    \cline{1-18}
   &   &   &\multicolumn{7}{c}{Example 3(a)}   &   &\multicolumn{7}{c}{Example 3(b)}
   \\    \cline{4-10}    \cline{12-18}
   &   &   &\multicolumn{3}{c}{Null hypotheses}    &    &\multicolumn{3}{c}{Alternatives}   &   &\multicolumn{3}{c}{Null hypotheses}    &    &\multicolumn{3}{c}{Alternatives}
   \\    \cline{4-6}    \cline{8-10}    \cline{12-14}    \cline{16-18}
   &   &   &$\rho=0.84$  &$\rho=0.82$  &$\rho=0.8$  &  &$\rho=0.75$  &$\rho=0.7$  &$\rho=0.65$  &   &$\rho=0.84$  &$\rho=0.82$  &$\rho=0.8$  &  &$\rho=0.75$  &$\rho=0.7$  &$\rho=0.65$
   \\    \cline{1-18}
$n=50$ &$p=2$ &$\gamma=0.5$   &0.0045 	&0.0150 	&0.0340 	&	&0.2125 	&0.4765 	&0.7400 	&	&0.0010 	&0.0045 	&0.0130 	&	&0.1245 	&0.3645 	&0.6510\\
      &      &$\gamma=1.0$    &0.0025 	&0.0105 	&0.0345 	&	&0.2345 	&0.5220 	&0.7825 	&	&0.0005 	&0.0070 	&0.0200 	&	&0.1610 	&0.4240 	&0.6945\\
      &      &$\gamma=1.5$    &0.0025 	&0.0100 	&0.0350 	&	&0.2550 	&0.5515 	&0.8030 	&	&0.0010 	&0.0085 	&0.0265 	&	&0.1890 	&0.4515 	&0.7085\\
      &      &$\gamma=2.0$    &0.0035 	&0.0095 	&0.0395 	&	&0.2615 	&0.5645 	&0.8115 	&	&0.0025 	&0.0095 	&0.0350 	&	&0.1935 	&0.4530 	&0.7025\\
    \cline{2-18}
      &$p=4$ &$\gamma=0.5$    &0.0000 	&0.0040 	&0.0205 	&	&0.2190 	&0.6080 	&0.8825 	&	&0.0000 	&0.0005 	&0.0065 	&	&0.0990 	&0.4035 	&0.7305\\
      &      &$\gamma=1.0$    &0.0005 	&0.0075 	&0.0320 	&	&0.2620 	&0.6255 	&0.8680 	&	&0.0035 	&0.0090 	&0.0290 	&	&0.1625 	&0.4195 	&0.6835\\
      &      &$\gamma=1.5$    &0.0040 	&0.0215 	&0.0545 	&	&0.2610 	&0.5750 	&0.8000 	&	&0.0110 	&0.0275 	&0.0535 	&	&0.1810 	&0.3750 	&0.5850\\
      &      &$\gamma=2.0$    &0.0125 	&0.0360 	&0.0680 	&	&0.2470 	&0.4930 	&0.7000 	&	&0.0195 	&0.0420 	&0.0725 	&	&0.1925 	&0.3600 	&0.5150\\
    \cline{2-18}
      &$p=6$ &$\gamma=0.5$    &0.0000 	&0.0005 	&0.0115 	&	&0.2220 	&0.6800 	&0.9280 	&	&0.0000 	&0.0000 	&0.0020 	&	&0.0635 	&0.3065 	&0.6315\\
      &      &$\gamma=1.0$    &0.0025 	&0.0110 	&0.0380 	&	&0.2745 	&0.6260 	&0.8685 	&	&0.0085 	&0.0205 	&0.0400 	&	&0.1605 	&0.3640 	&0.5855\\
      &      &$\gamma=1.5$    &0.0220 	&0.0490 	&0.0905 	&	&0.2900 	&0.5255 	&0.7145 	&	&0.0490 	&0.0730 	&0.1035 	&	&0.2220 	&0.3720 	&0.5205\\
      &      &$\gamma=2.0$    &0.0550 	&0.1120 	&0.1650 	&	&0.3290 	&0.4895 	&0.6305 	&	&0.0985 	&0.1320 	&0.1735 	&	&0.2865 	&0.4115 	&0.5230\\
    \cline{1-18}
$n=100$ &$p=2$ &$\gamma=0.5$  &0    	&0.0095 	&0.0475 	&	&0.3720 	&0.8010 	&0.9675 &	&0      	&0.0030 	&0.0185 	&	&0.2470 	&0.7190 	&0.9480\\
      &      &$\gamma=1.0$    &0    	&0.0095 	&0.0450 	&	&0.4145 	&0.8495 	&0.9815 &	&0      	&0.0040 	&0.0175 	&	&0.2945 	&0.7760 	&0.9610\\
      &      &$\gamma=1.5$    &0.0005 	&0.0095 	&0.0470 	&	&0.4500 	&0.8805 	&0.9880 &	&0.0005 	&0.0030 	&0.0240 	&	&0.3280 	&0.7835 	&0.9555\\
      &      &$\gamma=2.0$    &0.0010 	&0.0080 	&0.0525 	&	&0.4700 	&0.8870 	&0.9885 &	&0.0005 	&0.0050 	&0.0295 	&	&0.3360 	&0.7815 	&0.9490\\
    \cline{2-18}
      &$p=4$ &$\gamma=0.5$    &0    	&0.0015 	&0.0240 	&	&0.5130 	&0.9500 	&0.9985 &	&0      	&0.0005 	&0.0120 	&	&0.3385 	&0.8795 	&0.9960\\
      &      &$\gamma=1.0$    &0    	&0.0015 	&0.0305 	&	&0.5525 	&0.9485 	&0.9975 &	&0      	&0.0060 	&0.0265 	&	&0.3345 	&0.8085 	&0.9735\\
      &      &$\gamma=1.5$    &0    	&0.0070 	&0.0455 	&	&0.5010 	&0.9160 	&0.9920 &	&0.0030 	&0.0140 	&0.0430 	&	&0.3050 	&0.6680 	&0.9035\\
      &      &$\gamma=2.0$    &0.0005 	&0.0110 	&0.0510 	&	&0.4380 	&0.8470 	&0.9690 &	&0.0055 	&0.0205 	&0.0575 	&	&0.2810 	&0.5945 	&0.8300\\
    \cline{2-18}
      &$p=6$ &$\gamma=0.5$    &0    	&0.0005 	&0.0115 	&	&0.5795 	&0.9810 	&1 	    &	&0   	    &0      	&0.0030 	&	&0.2455 	&0.8430 	&0.9860\\
      &      &$\gamma=1.0$    &0.0005 	&0.0015 	&0.0255 	&	&0.5500 	&0.9605 	&0.9970 &	&0.0010 	&0.0080 	&0.0410 	&	&0.2565 	&0.6545 	&0.9020\\
      &      &$\gamma=1.5$    &0.0015 	&0.0155 	&0.0585 	&	&0.4400 	&0.8215 	&0.9585 &	&0.0130 	&0.0390 	&0.0750 	&	&0.2545 	&0.5145 	&0.7345\\
      &      &$\gamma=2.0$    &0.0100 	&0.0340 	&0.0895 	&	&0.3690 	&0.6545 	&0.8310 &	&0.0325 	&0.0595 	&0.0975 	&	&0.2705 	&0.4670 	&0.6460	\\
    \cline{1-18}	
$n=200$ &$p=2$ &$\gamma=0.5$  &0    	&0.0110 	&0.0570 	&	&0.6465 	&0.9795 	&1    	&	&0   	    &0.0025 	&0.0230 	&	&0.5350 	&0.9655 	&1\\
      &      &$\gamma=1.0$    &0 	    &0.0070 	&0.0585 	&	&0.7040 	&0.9895 	&1   	&	&0   	    &0.0025 	&0.0240 	&	&0.5930 	&0.9785 	&1\\
      &      &$\gamma=1.5$    &0    	&0.0050 	&0.0600 	&	&0.7510 	&0.9920 	&1      &	&0   	    &0.0030 	&0.0280 	&	&0.6195 	&0.9785 	&0.9990\\
      &      &$\gamma=2.0$    &0    	&0.0040 	&0.0615 	&	&0.7685 	&0.9920 	&1   	&	&0   	    &0.0040 	&0.0340 	&	&0.6095 	&0.9740 	&0.9990\\
    \cline{2-18}
      &$p=4$ &$\gamma=0.5$    &0    	&0.0010 	&0.0300 	&	&0.8545 	&0.9995 	&1   	&	&0   	    &0   	    &0.0085 	&	&0.7895 	&1   	    &1\\
      &      &$\gamma=1.0$    &0    	&0.0010 	&0.0325 	&	&0.8970 	&1      	&1   	&	&0   	    &0   	    &0.0165 	&	&0.6515 	&0.9895 	&1\\
      &      &$\gamma=1.5$    &0    	&0.0015 	&0.0390 	&	&0.8490 	&1      	&1   	&	&0   	    &0.0025 	&0.0295 	&	&0.4880 	&0.9400 	&0.9975\\
      &      &$\gamma=2.0$    &0    	&0.0020 	&0.0475 	&	&0.7800 	&0.9955 	&1   	&	&0   	    &0.0065 	&0.0370 	&	&0.4205 	&0.8715 	&0.9835\\
    \cline{2-18}
      &$p=6$ &$\gamma=0.5$    &0    	&0.0005 	&0.0165 	&	&0.9315 	&1 	        &1   	&	&0   	    &0   	    &0.0020 	&	&0.7040 	&0.9985 	&1\\
      &      &$\gamma=1.0$    &0    	&0.0010 	&0.0315 	&	&0.9080 	&1      	&1   	&	&0   	    &0.0030 	&0.0240 	&	&0.4515 	&0.9160 	&0.9970\\
      &      &$\gamma=1.5$    &0    	&0.0040 	&0.0490 	&	&0.7530 	&0.9935 	&1   	&	&0.0010 	&0.0165 	&0.0505 	&	&0.3425 	&0.7420 	&0.9300\\
      &      &$\gamma=2.0$    &0.0010 	&0.0105 	&0.0585 	&	&0.5825 	&0.9380 	&0.9935 &	&0.0045 	&0.0250 	&0.0740 	&	&0.3235 	&0.6530 	&0.8545\\
    \cline{1-18}
$n=300$ &$p=2$ &$\gamma=0.5$  &0    	&0.0010 	&0.0515 	&	&0.8065 	&0.9975 	&1   	&	&0   	    &0.0010 	&0.0195 	&	&0.7030 	&0.9965 	&1\\
      &      &$\gamma=1.0$    &0    	&0.0005 	&0.0500 	&	&0.8530 	&1      	&1   	&	&0   	    &0.0005 	&0.0235 	&	&0.7730 	&0.9990 	&1\\
      &      &$\gamma=1.5$    &0    	&0.0005 	&0.0425 	&	&0.8915 	&1      	&1   	&	&0   	    &0.0005 	&0.0290 	&	&0.7775 	&0.9985 	&1\\
      &      &$\gamma=2.0$    &0    	&0 	        &0.0460 	&	&0.8990 	&1      	&1   	&	&0   	    &0.0005 	&0.0310 	&	&0.7730 	&0.9985 	&1\\
    \cline{2-18}
      &$p=4$ &$\gamma=0.5$    &0    	&0      	&0.0400 	&	&0.9635 	&1      	&1   	&	&0   	    &0   	    &0.0125 	&	&0.9490 	&1   	    &1\\
      &      &$\gamma=1.0$    &0    	&0      	&0.0435 	&	&0.9810 	&1      	&1   	&	&0   	    &0.0005 	&0.0185 	&	&0.8230 	&1   	    &1\\
      &      &$\gamma=1.5$    &0    	&0      	&0.0505 	&	&0.9680 	&1      	&1   	&	&0   	    &0.0020 	&0.0255 	&	&0.6330 	&0.9870 	&1\\
      &      &$\gamma=2.0$    &0    	&0      	&0.0555 	&	&0.9275 	&1      	&1   	&	&0.0005 	&0.0040 	&0.0315 	&	&0.5445 	&0.9585 	&0.9995\\
    \cline{2-18}
      &$p=6$ &$\gamma=0.5$    &0    	&0      	&0.0185 	&	&0.9950 	&1      	&1   	&	&0   	    &0   	    &0.0050 	&	&0.9150 	&1   	    &1\\
      &      &$\gamma=1.0$    &0    	&0      	&0.0310 	&	&0.9895 	&1      	&1   	&	&0   	    &0.0015 	&0.0290 	&	&0.6050 	&0.9850 	&1\\
      &      &$\gamma=1.5$    &0    	&0.0020 	&0.0495 	&	&0.9115 	&0.9995 	&1   	&	&0.0015 	&0.0090 	&0.0435 	&	&0.4310 	&0.8720 	&0.9875\\
      &      &$\gamma=2.0$    &0    	&0.0085 	&0.0580 	&	&0.7655 	&0.9945 	&1   	&	&0.0020 	&0.0165 	&0.0545 	&	&0.3875 	&0.7920 	&0.9515\\
    \cline{1-18}
  \end{tabular}
  }
\end{table}%

\newpage
\begin{table}[p]
  \centering
  \caption{Values of $\Delta^{(\mathcal H)}_{w,Z}$ for the benchmark configuration in Example 2(a) based on the random approximation (RA) and numerical integration (NI).} \label{tableH1}
  \renewcommand\arraystretch{1.2}
  \scalebox{0.85}{
    \begin{tabular}{cccccccccccccc}
    \cline{1-10}
                     &               &\multicolumn{2}{c}{$p=2$} & &\multicolumn{2}{c}{$p=4$} & &\multicolumn{2}{c}{$p=6$} \\
                     \cline{3-4}    \cline{6-7}    \cline{9-10}
     Weight function &                &RA           &NI         & &RA           &NI         & &RA           &NI  \\
    \cline{1-10}
     Stable density  &$\gamma=0.5$    &0.216287 	&0.216954 	& &0.177086 	&0.182100 	& &0.154451 	&0.150669\\
                     &$\gamma=1.0$    &0.312581 	&0.315284 	& &0.170617 	&0.171631 	& &0.097647 	&0.098159\\
                     &$\gamma=1.5$    &0.320700 	&0.324544 	& &0.107838 	&0.108673 	& &0.034544 	&0.034622\\
                     &$\gamma=2.0$    &0.314935 	&0.319257 	& &0.076042 	&0.076714 	& &0.015729 	&0.015833\\
    \cline{1-10}
     Laplace density &$\gamma=0.1$    &0.160431 	&0.160761 	& &0.167135 	&0.167664 	& &0.163473 	&0.161230\\
                     &$\gamma=0.25$   &0.307405 	&0.308412 	& &0.290496 	&0.291522 	& &0.266015 	&0.264073\\
                     &$\gamma=1.0$    &0.388005 	&0.391292 	& &0.221240 	&0.222430 	& &0.141637 	&0.141809\\
                     &$\gamma=4.0$    &0.122471 	&0.124460 	& &0.014131 	&0.014279 	& &0.001981   	&0.002002\\
    \cline{1-10}
  \end{tabular}
  }
\end{table}%
%

\newpage
\begin{table}[p]
  \centering
  \caption{p--values of $T^{(\mathcal I)}_{n,S}$, $T^{(\mathcal I)}_{n,L}$, and $T^{(\mathcal I)}_{n,E}$ with $\gamma=1.0$ and different values of $\Delta$ for the analysis of ACTG175 data set.} \label{tableRealInde}
  \renewcommand\arraystretch{1.2}
  \scalebox{0.85}{
    \begin{tabular}{ccccccccccccccccccccc}
    \cline{1-5}
        &$\Delta^{(\mathcal I)}_{n,w}$  &$\Delta$  &${\sqrt{n}(\Delta^{(\mathcal I)}_{n,w}-\Delta)}/{\sigma_{n,w}^{(\mathcal I)}}$  &p--value\\
    \cline{1-5}
     $T^{(\mathcal I)}_{n,S}$  &$0.000525$   &$0.0003$    &1.0533 	&0.8539\\
                               &             &$0.0006$    &-0.3536 	&0.3681\\
                               &             &$0.0009$    &-1.7605 	&0.0392\\
                               &             &$0.0012$    &-3.1674  &0.0008\\
    \cline{1-5}
     $T^{(\mathcal I)}_{n,L}$  &$0.000696$   &$0.0003$    &1.3636 	&0.9137\\
                               &             &$0.0006$    &0.3312 	&0.6298\\
                               &             &$0.0009$    &-0.7012 	&0.2416\\
                               &             &$0.0012$    &-1.7336  &0.0415\\
    \cline{1-5}
     $T^{(\mathcal I)}_{n,E}$  &$0.011890$  &$0.005$   	&1.4754 	&0.9299\\
                               &            &$0.010$    &0.4047 	&0.6572\\
                               &            &$0.015$    &-0.6660 	&0.2527\\
                               &            &$0.020$    &-1.7367 	&0.0412\\
    \cline{1-5}
  \end{tabular}
  }
\end{table}%

\newpage

\newpage

\newpage

\newpage

\newpage

\newpage

\newpage

\end{document}